\documentclass[twoside]{article}

\usepackage{amssymb}
\usepackage{amsmath}
\usepackage{amsfonts}
\usepackage{ascmac}
\usepackage{qic,epsfig}

\usepackage{color}
\usepackage{lscape}
\usepackage{comment}
\usepackage{multirow}
\usepackage{cancel}
\usepackage{ulem}
\usepackage{algorithm}
\newtheorem{lem}[theorem]{Lemma}
\newtheorem{defi}[theorem]{Definition}
\newtheorem{cor}[theorem]{Corollary}

\newtheorem{proposition}[theorem]{Proposition}
\newcommand{\bra}[1]{\langle #1\rvert}
\newcommand{\ket}[1]{\lvert #1\rangle}
\newcommand{\norm}[1]{\lVert #1\rVert}
\newcommand{\Norm}[1]{\left\lVert #1\right\rVert}

\newcommand{\ketbra}[2]{\left \lvert #1\rangle\langle #2 \right \rvert}

\newcommand{\gaiseki}{\ketbra}

\newcommand{\im}{\mathrm{{\bf i}}}   %%
\newcommand{\e}{\mathrm{e}}          %%
\newcommand{\abs}[1]{\left|#1\right|} %

\newcommand{\C}{\mathbb{C}} %
\newcommand{\Z}{\mathbb{Z}}
\newcommand{\tenti}{\mathrm{T}}

\newcommand{\dz}{\mathrm{d}z}
\newcommand{\dtheta}{\mathrm{d}\theta}

\newcommand{\poly}{\textrm{poly}}
\newcommand{\diag}{\textrm{diag}}

\newcommand{\qstate}[1]{\frac{#1}{\norm{#1}}}
\newcommand{\qstateinline}[1]{#1/\norm{#1}}
\renewcommand{\vec}[1]{\mbox{\boldmath $#1$}} %%
\renewcommand{\O}{O}
\newcommand{\qed}{\hfill $\Box$}
\newcommand{\myproof}[1]{{\bf Proof.} #1 \hfill $\Box$}% 変更．オリジナル→\newcommand{\proof}[1]{{\bf Proof.} #1 $\Box$.}
\renewcommand{\proof}[1]{{\bf Proof.} #1 \hfill $\Box$}% 変更．オリジナル→\newcommand{\proof}[1]{{\bf Proof.} #1 $\Box$.}

\begin{document}
\setlength{\textheight}{8.0truein}    %FOR 2ND PAGE ONWARDS

%\runninghead{Title  $\ldots$}     本来の
%            {Author(s) $\ldots$}
\runninghead
{Quantum algorithm for matrix functions by cauchy's integral formula}
{S. Takahira, A. Ohashi, T. Sogabe, T.S. Usuda}

\normalsize\textlineskip
\thispagestyle{empty}
\setcounter{page}{1}

%\copyrightheading{Vol.}{No.}{Year}{Page Nos.}
\copyrightheading{0}{0}{2003}{000--000}

\vspace*{0.88truein}

\alphfootnote

\fpage{1}

\centerline{\bf
%%%%%%%%%%%%%%%%%%%%%
%Put in titiles here
%%%%%%%%%%%%%%%%%%%%%
QUANTUM ALGORITHM FOR MATRIX FUNCTIONS}
\vspace*{0.035truein}
\centerline{\bf BY CAUCHY'S INTEGRAL FORMULA}
\vspace*{0.37truein}
\centerline{\footnotesize
%%%%%%%%%%%%%%%%%%%%%%%%%%%%%%%%%%%%
%put authors' name and address here
%%%%%%%%%%%%%%%%%%%%%%%%%%%%%%%%%%%%
SOUICHI TAKAHIRA}
\vspace*{0.015truein}
\centerline{\footnotesize\it
Graduate School of Information Science and Technology, Aichi Prefectural University}
\baselineskip=10pt
\centerline{\footnotesize\it
1522-3 Ibaragabasama, Nagakute, Aichi, 480-1198, Japan}
\vspace*{10pt}
\centerline{\footnotesize
ASUKA OHASHI}
\vspace*{0.015truein}
\centerline{\footnotesize\it
College of Science and Engineering, Ritsumeikan University}
\baselineskip=10pt
\centerline{\footnotesize\it
1-1-1 Noji-Higashi, Kusatsu, Shiga, 525-8577, Japan}
\vspace*{10pt}
\centerline{\footnotesize
TOMOHIRO SOGABE}
\vspace*{0.015truein}
\centerline{\footnotesize\it
Graduate School of Engineering, Nagoya University}
\baselineskip=10pt
\centerline{\footnotesize\it
Furo-cho, Chikusa, Nagoya, 464-8603, Japan}
\vspace*{10pt}
\centerline{\footnotesize
TSUYOSHI SASAKI USUDA}
\vspace*{0.015truein}
\centerline{\footnotesize\it
Graduate School of Information Science and Technology, Aichi Prefectural University}
\baselineskip=10pt
\centerline{\footnotesize\it
1522-3 Ibaragabasama, Nagakute, Aichi, 480-1198, Japan}
\vspace*{0.225truein}
\publisher{(received date)}{(revised date)}

\vspace*{0.21truein}

%% \abstracts{first paragraph}{second paragraph}{third paragraph}
%% If there is only one paragraph, just keep the second and third empty
%% like the following one
\abstracts{
For matrix $A$, vector $\vec{b}$ and function $f$,
the computation of vector $f(A)\vec{b}$ arises in many scientific computing applications.
We consider the problem of obtaining quantum state $\ket{f}$ corresponding to vector $f(A)\vec{b}$.
There is a quantum algorithm to compute state $\ket{f}$ using eigenvalue estimation
that uses phase estimation and Hamiltonian simulation $\e^{\im A t}$.
However, the algorithm based on eigenvalue estimation needs $\poly(1/\epsilon)$ runtime,
where $\epsilon$ is the desired accuracy of the output state.
Moreover, if matrix $A$ is not Hermitian,
$\e^{\im A t}$ is not unitary and we cannot run eigenvalue estimation.
In this paper, we propose a quantum algorithm that uses
Cauchy's integral formula and the trapezoidal rule
as an approach that avoids eigenvalue estimation.
We show that the runtime of the algorithm is $\poly(\log(1/\epsilon))$ %$O(\log(1/\epsilon))$
and the algorithm outputs state $\ket{f}$ even if $A$ is not Hermitian.
}{}{}

\vspace*{10pt}

\vspace*{3pt}
\communicate{to be filled by the Editorial}

\vspace*{1pt}\textlineskip    %) USE THIS MEASUREMENT WHEN THERE IS
   %) A SECTION HEADING
%\vspace*{-0.5pt}
%\noindent
%%%%%%%%%%%%%%%%%%%%%%%%%%%%%%%%
%put the text of the paper here
%%%%%%%%%%%%%%%%%%%%%%%%%%%%%%%%

%\tableofcontents

% \input{v36_Intro.tex}
\section{Introduction}\label{Introdution}
Many quantum algorithms have been proposed.
For example, Shor's factoring algorithm \cite{Shor:1994} and Grover's search algorithm \cite{Grover:1996}.
Among them, the quantum algorithm for linear system
$A\vec{x} = \vec{b}$ proposed by Harrow, Hassidim and Lloyd \cite{HHL:2009},
in particular, has wide applications.
The authors' quantum algorithm, which is called the HHL algorithm,
can output quantum state
$\ket{x} = \qstateinline{\sum_{i=0}^{N-1}x^{[i]}\ket{i}}$
corresponding to solution $\vec{x} = (x^{[0]},x^{[1]},\dots,x^{[N-1]})^\tenti$
with $O(\log N)$, where $N$ is the size of the solution.
Because linear systems often arise in scientific computing and the HHL algorithm is very efficient for size $N$,
this algorithm is used as a subroutine in many quantum algorithms,
for example,  \cite{WBL:2012, CJS:2013, RML:2015, MP:2016, BCOW:2017}.

The HHL algorithm \cite{HHL:2009} is based on eigenvalue estimation.
The procedure is as follows:
First, we determine eigenvalues $\lambda_j$ of matrix $A$ for the eigenvectors
using a Hamiltonian simulation algorithm and the phase estimation algorithm.
%=================================================================================================================
Second, we perform a rotation based on the value of $\lambda_j$ to an ancilla qubit.
Performing this rotation, the ancilla qubit becomes $C\lambda_j^{-1}\ket{0} + \sqrt{1 - |C\lambda_j^{-1}|^2}\ket{1}$,
where $C$ is a normalizing constant.
Third, we apply the inverse of the phase estimation algorithm.
Finally, we measure the ancilla qubit. If the result is $0$,
the eigenvectors are multiplied by the inverse of their corresponding eigenvalues $\lambda_j$.
%=================================================================================================================
Thus, we obtain state $\ket{x}$.
If the result is $1$, then the algorithm fails.
Therefore, we use the amplitude amplification \cite{BHMT:2002} to boost the success probability before the measurement.
Moreover, by replacing
%=================================================================================================================
the rotation operator in the second step with a rotation operator
to obtain state $Cf(\lambda_j)\ket{0} + \sqrt{1 - |Cf(\lambda_j)|^2}\ket{1}$,
%=================================================================================================================
Harrow, Hassidim and Lloyd mentioned \cite{HHL:2009} that the HHL algorithm can be generalized to
a quantum algorithm that computes quantum state
\begin{align}
  \ket{f} := \frac{f(A)\ket{b}}{\norm{f(A)\ket{b}}}
  % \ket{f} := \frac{ \sum_{i=0}^{N-1}(f(A)\vec{b})^{[i]}\ket{i} }{\norm{ \sum_{i=0}^{N-1}(f(A)\vec{b})^{[i]}\ket{i} } }
  ,
  \label{def:matrixfunctionstate}
\end{align}
where $f(A) \in \C^{N \times N}$ denotes a matrix function
%=================================================================================
%=================================================================================
and $\ket{b} = \qstateinline{\sum_{i=0}^{N-1} b^{[i]}\ket{i}}$ is the quantum state corresponding to vector
$\vec{b} = (b^{[0]}, b^{[1]}, \dots, b^{[N-1]})^\tenti$.
%=================================================================================
Matrix function
$f(A)$ is defined as
\begin{align}
  f(A) := a_0 I_N + a_1 A + a_2 A^2 + \cdots = \sum_{j=0}^\infty a_j A^j,
  \label{eq:f(A)}
\end{align}
for function $f(z) = a_0 + a_1z^1 + a_2z^2 + \cdots \ (a_j, z \in \C)$,
where $I_N$ denotes the $N\times N$ identity matrix.

The computation of matrix function $f(A)$ arises in scientific computing.
For example, differential equations and exponential integrators \cite{Higham}.
In some applications, there is the case in which matrix-vector product $f(A)\vec{b}$ is
computed for vector $\vec{b}$ \cite{Higham}.
Therefore, in terms of applications, it is also important to compute state $\ket{f}$.

However, the quantum algorithm \cite{HHL:2009} based on eigenvalue estimation needs
$\poly(1/\epsilon)$ %$\O(1/\epsilon)$
runtime, where $\epsilon$ is the desired accuracy of the output state,
because the method uses the phase estimation algorithm.
Additionally, matrix $A$ must be Hermitian because the quantum algorithm uses a Hamiltonian simulation algorithm.

In this study, we propose a simple quantum algorithm to compute state $\ket{f}$ with
$\poly(\log(1/\epsilon))$ %$\O(\log(1/\epsilon))$
runtime even if matrix $A$ is not Hermitian.
The idea is as follows:
Using Cauchy's integral formula and the trapezoidal rule,
the matrix function can be represented as a weighted sum of solutions of linear systems (see e.g., \cite{Higham, Trefethen}).
Considering this fact,
we can replace the problem of obtaining state $\ket{f}$ with the problem of obtaining
the quantum state corresponding to the solution of a block diagonal linear system.
The HHL algorithm can output the state corresponding to the solution,
even if the coefficient matrix is not Hermitian.
Furthermore, there is an improved version \cite{CKS:2015} of the HHL algorithm that
outputs the state corresponding to the solution with $\poly(\log(1/\epsilon))$ % $\O(\log(1/\epsilon))$
runtime. Thus, we can obtain state $\ket{f}$ with $\poly(\log(1/\epsilon))$ %$\O(\log(1/\epsilon))$
runtime even if matrix $A$ is not Hermitian.
In this paper, we describe the proposed quantum algorithm to compute state $\ket{f}$
and analyze the runtime, error and success probability of the proposed quantum algorithm.

Quantum algorithms to obtain state $\ket{f}$ with $\poly(\log(1/\epsilon))$ runtime
have already been shown in \cite{AGGW:2017, SBJ:2018, GSLW:2018}.
The difference between the proposed quantum algorithm and those quantum algorithms is
whether the matrix decomposition of matrix function $f(A)$ is used.
Additionally, we can apply the proposed quantum algorithm to the case in which matrix $A$ is not Hermitian.
The method based on eigenvalue decomposition in \cite{AGGW:2017,  SBJ:2018} uses the property that matrix $A$ is Hermitian.
The method based on singular value decomposition in \cite{GSLW:2018} is applicable
when matrix function $f(A)$ is defined as $f(A) = Uf(\Sigma) V^\dag$
for singular value decomposition $A = U\Sigma V^\dag$ of matrix $A$.
By contrast, our method does not set such a condition.

\subsection{Problem statement}\label{ProblemStatement}
We define the problem of obtaining state $\ket{f}$ formally.
For simplicity, we assume that $N=2^n$, where $n$ is a positive integer.
For matrix $A \in \C^{N \times N}$ and vector $\vec{b} \in \C^N$,
as in \cite{HHL:2009}, we make the following assumptions.

We assume that matrix $A \in \C^{N \times N}$ satisfies $\norm{A} \le 1$
and has at most $d$ nonzero elements in any row or column,
where $\norm{A}$ represents the spectral norm of matrix $A$.
Matrix $A$ is called a $d$-\textit{sparse} matrix
when the maximum number of nonzero elements in any row or column is $d$.
For $d$-sparse matrix $A$,
we assume there is oracle $\mathcal{P}_A$, which
consists of oracle $O_A$ and oracle $O_{\nu}$,
where $O_A$ is the unitary operator that returns elements for given positions
and $O_{\nu}$ is the unitary operator that returns the positions of the nonzero elements.
Specifically,
oracle $O_A$ is the unitary operator such that
\begin{align}
  O_A\ket{i,j,z} = \ket{i,j,z \oplus A_{ij}},
  \label{oracle_A}
\end{align}
for $i,j \in \{0,1,\dots,N-1\}$,
where $A_{ij}$ denotes the binary representation of the $(i,j)$ element of matrix $A$
and $\oplus$ denotes the bitwise XOR operation.
Oracle $O_\nu$ is the unitary operator such that
\begin{align}
 O_\nu \ket{j, \ell} = \ket{j, \nu(j, \ell)}, \qquad
 \label{oracle_nu}
\end{align}
for $j \in \{0,1,\dots,N-1\}$ and $\ell \in \{0,1,\dots,d-1\}$,
where $\nu(j,\ell)$ is a function that returns the row index of
the $\ell$-th nonzero element in the $j$-th column.
For more details on these oracles, see, for example, \cite{BC:2012}.
For given vector
$\vec{b} = (b^{[0]}, b^{[1]}, \dots, b^{[N-1]})^\tenti \in \C^N$,
we assume there is oracle $\mathcal{P}_{\vec{b}}$ that generates state
$\ket{b} = \qstateinline{\sum_{i=0}^{N-1}b^{[i]}\ket{i}}$
corresponding to vector $\vec{b}$ with $O(\log N)$ runtime;
that is, we assume that there is unitary operator $\mathcal{P}_{\vec{b}}$ such that
\begin{align}
  \mathcal{P}_{\vec{b}}\ket{0^n} = \ket{b}.
  \label{oracle_Pb}
\end{align}

Using the oracles $\mathcal{P}_A$ and $\mathcal{P}_{\vec{b}}$,
we define the problem for matrix functions as follows:

\begin{defi}[Quantum Matrix Function Problem]\label{QMFP}
  Let $A \in \C^{N \times N}$ be an $N\times N$ $d$-sparse matrix that satisfies $\norm{A} \le 1$
  and let $\vec{b} \in \C^N$ be an $N$-dimensional complex vector.
  Suppose that there is oracle $\mathcal{P}_A$,
  which consists of oracles $O_A$ and $O_{\nu}$ in Eqs.~\eqref{oracle_A} and \eqref{oracle_nu}, respectively,
  and oracle $\mathcal{P}_{\vec{b}}$ in Eq.~\eqref{oracle_Pb}.
  For matrix function $f(A)$ in Eq.~\eqref{eq:f(A)} and vector $\vec{b}$,
  we define quantum state $\ket{f}$ as
  %=============================================================================
  % $\ket{f} := \qstateinline{ \sum_{i=0}^{N-1}(f(A)\vec{b})^{[i]}\ket{i}}$,
  % where ($f(A)\vec{b})^{[i]}$ denotes the $i$-th element of vector $f(A)\vec{b}$
  $\ket{f} = f(A)\ket{b}/\norm{f(A)\ket{b}}$.
  %=============================================================================
  Then, for some positive constant $\epsilon$,
  the problem is to output quantum state $\ket{\tilde{f}}$ such that
  \begin{align}
    \Norm{
      \ket{f} - \ket{\tilde{f}}
    }
    \le \epsilon,
  \end{align}
  with a probability of at least $1/2$,
  where $0 \le \epsilon \le 1/2$.
\end{defi}

\subsection{Main result}
In this study, for the problem defined by Definition \ref{QMFP}, we obtain the following result.

\begin{theorem}[Main Result]\label{MainResult}
  Suppose that $R > 1$, and $f(z)$ is an analytic function on disk $|z| \le R$.
  Let $B$ be the maximum value of $|f(z)|$ on disk $|z| \le R$ and
  let $\beta$ be a real number such that $1 < \beta < R$.
  Then the problem defined by Definition \ref{QMFP} can be solved using
  \begin{align}
    O\left( \frac{d\kappa'^2}{F(1-r)} \log^2 \left( \frac{d\kappa'}{F}\frac{1}{\epsilon} \right) \right)
    \text{ queries to $\mathcal{P}_A$ and }
    O\left( \frac{\kappa'}{F(1-r)}\log\left( \frac{d\kappa'}{F}\frac{1}{\epsilon} \right)  \right)
    \text{ uses of $\mathcal{P}_{\vec{b}}$},
  \end{align}
  with gate complexity
  \begin{align}
    & O\Biggl(
      \frac{d\kappa'^2}{F(1-r)} \log^2 \left(  \frac{d\kappa'}{F}\frac{1}{\epsilon} \right)
      \left[
      \log(N)  + \log(\gamma) +
      \log^{\frac{5}{2}}\left( \frac{d \kappa'}{F}\frac{1}{\epsilon} \right)
      \right]
      \notag \\
    &\qquad\qquad  +
      \frac{1}{F(1-r)^2}\log\left(\frac{1}{F(1-r)}\frac{1}{\epsilon} \right)
      +
      \frac{1}{F(1-r)}\log(\gamma) \log\left(\frac{1}{1-r}\right)
    \Biggr)
    ,
    \label{theorem:gate}
  \end{align}
  where
  $r := \beta/R, \ \kappa' = 1/(1-\beta^{-1}), \ \gamma = \max\{\kappa', 1/(1-r)\}$, and $F = \norm{f(A)\ket{b}}/(B\kappa')$.
\end{theorem}

\begin{figure}[htbp]
  \centering
  \includegraphics[clip,width=12.0cm]{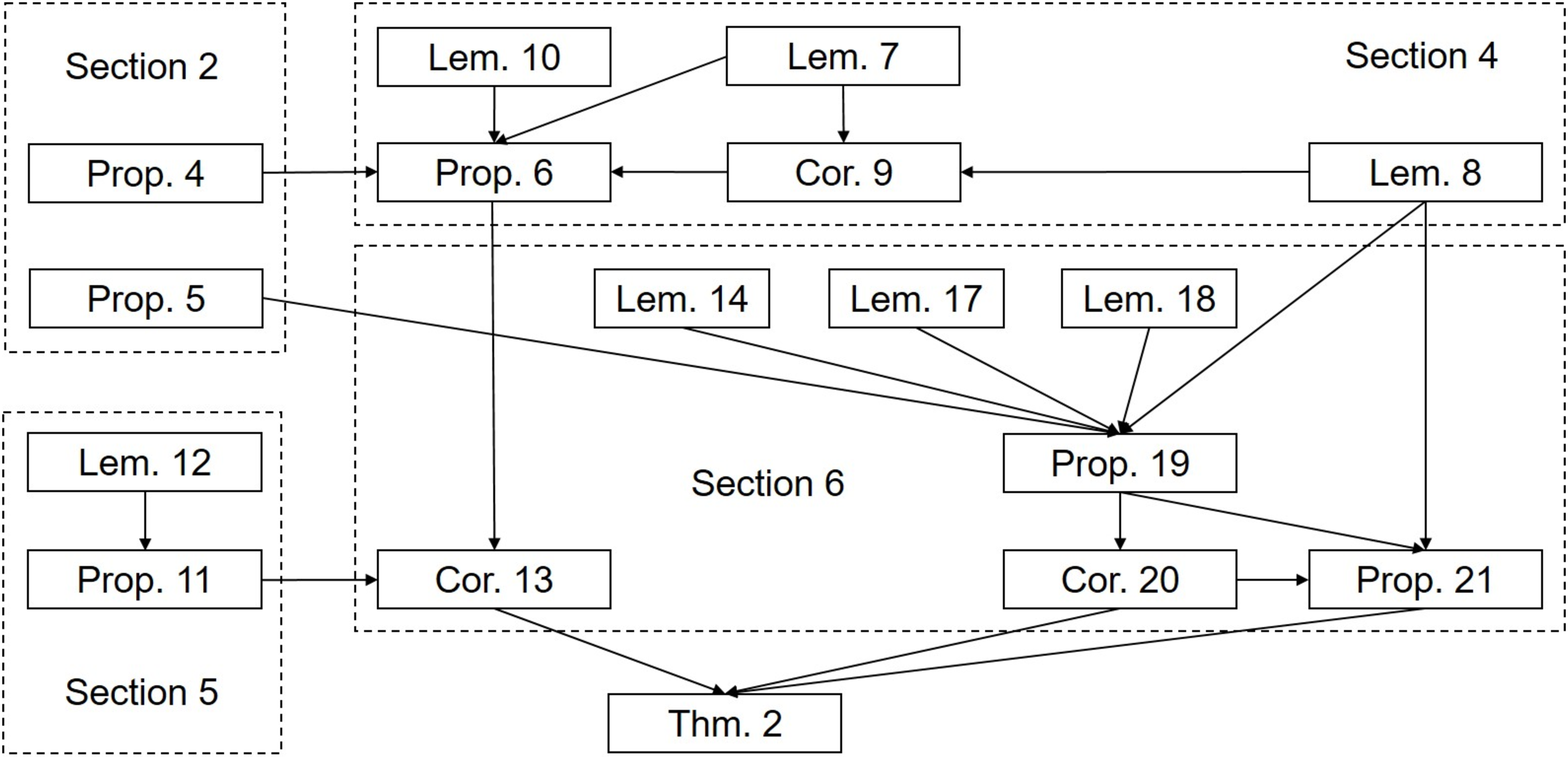}
  \fcaption{The roadmap to obtain Theorem \ref{MainResult}}
  \label{fig:map}
\end{figure}

The proof of this theorem is in Section \ref{subsec:ProofOfTheo}.
Before the proof, we present some propositions, lemmas, and corollaries.
The roadmap to obtain  Theorem \ref{MainResult} is shown in Fig. \ref{fig:map}.

This paper is organized as follows:
In Section \ref{Preliminaries},
we present the result of \cite{CKS:2015},
which shows the improved version of the HHL algorithm,
and the approximation of the matrix function.
In Section \ref{QuantAlgo},
we propose a quantum algorithm to compute quantum state $\ket{f_M}$,
which approximates state $\ket{f}$.
The proposed quantum algorithm uses two subroutines.
The first subroutine is the HHL algorithm to solve the block diagonal linear system.
We consider applying the HHL algorithm to the block diagonal linear system in Section \ref{sec:LS}.
The second subroutine is a unitary operator for multiplying weights.
In Section \ref{sec:Uf}, we consider this unitary operator.
In Section \ref{ErrorRuntime}, we analyze the runtime, error and
success probability of the proposed quantum algorithm.
Then, we prove the main theorem using the analysis.
Finally, we conclude this paper in Section \ref{Conclusion}.

Throughout this paper,
we suppose that complex function $f(z)$ satisfies the assumptions in Theorem \ref{MainResult}.
Furthermore, let $B, \beta$ and $r$ be defined as in Theorem \ref{MainResult}.
Additionally, for square matrix $X$ and vector $\vec{v}$,
$\norm{X}$ and $\norm{\vec{v}}$ represents the spectral norm of matrix $X$
and the $\ell^2$-norm of vector $\vec{v}$, respectively.
Moreover, $v^{[i]}$ represents the $i$-th element of vector $\vec{v}$.

%-------------------------------------------------------------------------------
%\input{v36_Pleliminaries.tex}
\section{Preliminaries}\label{Preliminaries}
In this section, we present the background briefly.
In Section \ref{sec:HHL}, we describe the problem for linear systems and the result of the improved version \cite{CKS:2015} of the HHL algorithm.
In Section \ref{sec:Approx}, we describe the representation of matrix function $f(A)$ using Cauchy's integral formula
and the approximation using the trapezoidal rule.

\subsection{Improved version of the HHL algorithm}\label{sec:HHL}
We describe the problem, which is called the quantum linear systems problem (QLSP),
for linear system $A\vec{x}=\vec{b}$.
For the details of this problem and the results, see \cite{HHL:2009, CKS:2015, Primer}.

%------------------------------------------------------------------------------
\begin{defi}[Quantum Linear Systems Problem]\label{QLSP}
Let $A \in \C^{N \times N}$ be an $N\times N$ $d$-sparse Hermitian matrix that satisfies $\norm{A} \le 1$
and let $\vec{b} \in \C^N$ be an $N$-dimensional complex vector.
Suppose that there is oracle $\mathcal{P}_A$,
which consists of oracles $O_A$ and $O_{\nu}$ in Eqs.~\eqref{oracle_A} and \eqref{oracle_nu}, respectively,
and oracle $\mathcal{P}_{\vec{b}}$ in Eq.~\eqref{oracle_Pb}.
For linear system $A\vec{x}=\vec{b}$,
we define quantum state $\ket{x}$ as
%===============================================================================
$\ket{x}:= \qstateinline{\sum_{i=0}^{N-1}x^{[i]}\ket{i}}$, where $x^{[i]}$ is the $i$-th element of solution $\vec{x}$.
%===============================================================================
Then, for some positive constant $\epsilon$,
the problem is to output state $\ket{\tilde{x}}$ such that
\begin{align}
  \Bigl\lVert \ket{x} - \ket{\tilde{x}} \Bigr\rVert \le \epsilon,
\end{align}
with a probability of at least $1/2$,
where $0 \le \epsilon \le 1/2$.
\end{defi}
%------------------------------------------------------------------------------
Note that we can remove the condition that
matrix $A$ is Hermitian without loss of generality \cite{HHL:2009}.
To see this, consider the following linear system,
\begin{align}
  \begin{bmatrix}
    0 & A \\
    A^\dag & 0
  \end{bmatrix}
  \begin{bmatrix}
    0 \\ \vec{x}
  \end{bmatrix}
  =
  \begin{bmatrix}
    \vec{b} \\ 0
  \end{bmatrix},
  \label{extendedlinearsystems}
\end{align}
instead of linear system $A\vec{x} = \vec{b}$.
Then the solution of linear system \eqref{extendedlinearsystems} is essentially
solution $\vec{x}$ of linear system $A\vec{x}=\vec{b}$.
Furthermore, a state corresponding to $(0, \vec{x}^\tenti)^\tenti$ is also essentially quantum state $\ket{x}$.

For this problem, the following is known.

%------------------------------------------------------------------------------
\begin{proposition}\cite[Theorem 4]{CKS:2015}\label{ImprovedVersionOfHHL}
  The QLSP defined in Definition \ref{QLSP} can be solved using
  \begin{align}
   O\left( d\kappa_A^2\log^2 \left(\frac{d\kappa_A}{\epsilon}\right) \right)
   \text{ queries to oracle $\mathcal{P}_A$ and }
   O\left( \kappa_A\log\left(\frac{d\kappa_A}{\epsilon}\right) \right)
   \text{ uses of $\mathcal{P}_{\vec{b}}$},
  \end{align}
  with gate complexity
  \begin{align}
    O\left(
    d\kappa_A^2\log^2\left( \frac{d\kappa_A}{\epsilon} \right)
    \left[ \log N + \log^{\frac{5}{2}}\left(\frac{d\kappa_A}{\epsilon} \right) \right]
    \ \right),
  \end{align}
  where $\kappa_A := \norm{A}\norm{A^{-1}}$ is the condition number of matrix $A$.
\end{proposition}
%------------------------------------------------------------------------------
The proposed algorithm uses the improved version \cite{CKS:2015} of the HHL algorithm as a subroutine.
This proposition is used in the proof of Proposition \ref{prop:LS} for the subroutine.

\subsection{Approximation by Cauchy's integral theorem and the trapezoidal rule}\label{sec:Approx}
Let $\Gamma$ be a closed contour in the complex plane that encloses all eigenvalues of matrix $A$
and let $f$ be an analytic function on and inside $\Gamma$.
Then, using Cauchy's integral formula, matrix function $f(A)$ can be described as
\begin{align}
  f(A) = \frac{1}{2\pi\im}\int_\Gamma f(z)(zI_N - A)^{-1} \dz,
\end{align}
where $I_N$ denotes the $N\times N$ identity matrix.
As contour $\Gamma$, we can consider circle $z = \beta\e^{\im \theta} \ (0 \le \theta \le 2\pi)$
with center $0$ and radius $\beta$.
Therefore, matrix function $f(A)$ can be represented as
\begin{align}
  f(A) = \frac{1}{2\pi\im}\int_0^{2\pi} f(\beta\e^{\im \theta})(\beta\e^{\im \theta}I_N - A)^{-1} \im\beta\e^{\im\theta} \dtheta = \int_0^{2\pi} h(\theta) \dtheta,
  \label{integral}
\end{align}
where $h(\theta) = f(\beta\e^{\im \theta})(\beta\e^{\im \theta}I_N - A)^{-1} \beta\e^{\im\theta}/2\pi$.
Next, we construct approximation $f_M(A)$ of matrix function $f(A)$.
Let
\begin{align}
  \theta_k := \frac{2\pi}{M}k.
\end{align}
We consider $M$ points $\{ \beta\e^{\im\theta_k} \mid k = 0,1,\dots,M-1\}$
at regular intervals on the circle and apply the trapezoidal rule to integral \eqref{integral}.
Then, as we can see in the following, we have approximation $f_M(A)$.
\begin{align}
  f(A) = \sum_{k=0}^{M-1} \int_{\theta_k}^{\theta_{k+1}} h(\theta) \dtheta
  &\simeq \sum_{k=0}^{M-1} (\theta_{k+1}-\theta_k) \frac{h(\theta_k) + h(\theta_{k+1})}{2} \notag\\
  &= \frac{1}{M}\sum_{k=0}^{M-1} f(\beta\e^{\im \theta_k}) \beta\e^{\im\theta_k} (\beta\e^{\im \theta_k}I_N - A)^{-1} \notag\\
  &=: f_M(A).
  \label{fMA}
\end{align}
Hereafter, $M = 2^m$, where $m$ is a positive integer.
We describe the error in approximation $f_M(A)$.
Specifically, the following proposition holds.
The following proposition is used in the proof of Lemma \ref{Manddelta} with respect to
the error of the proposed quantum algorithm.

%------------------------------------------------------------------------------
\begin{proposition}\cite[Theorem 18.1]{Trefethen}\label{approxTheorem}
  For matrix function $f(A)$ in Eq.~\eqref{eq:f(A)} and approximation $f_M(A)$ in Eq.~\eqref{fMA},
  we have a bound
  \begin{align}
    \Norm{f(A) - f_M(A)}
    &\le
    \frac{B}{1 - \frac{\norm{A}}{R}}\left( \frac{1}{1 - (\frac{\norm{A}}{\beta})^M }\left(\frac{\norm{A}}{\beta}\right)^M +
    \frac{1}{1 - (\frac{\beta}{R})^M}\left(\frac{\beta}{R}\right)^M \right),
     \label{fAminusfMA}
  \end{align}
  where $B, \beta$, and $R$ are the positive real numbers defined as in Theorem \ref{MainResult}.
\end{proposition}
%------------------------------------------------------------------------------
\proof{
  In \cite{Trefethen}, the proof is not given.
  Therefore, we provide the proof.
  Because $\norm{A} \le 1 < \beta$,
  $\beta\e^{\im\theta_k}(\beta\e^{\im\theta_k}I_N - A)^{-1} = \sum_{\ell = 0}^\infty (A/(\beta\e^{\im\theta_k}))^\ell$ holds.
  Thus,
  \begin{align}
    f_M(A)
    &= \frac{1}{M} \sum_{k=0}^{M-1} f(\beta\e^{\im\theta_k}) \sum_{\ell = 0}^\infty \left(\frac{A}{\beta\e^{\im\theta_k}}\right)^\ell \notag\\
    &= \frac{1}{M} \sum_{k=0}^{M-1} \sum_{j=0}^\infty a_j (\beta\e^{\im\theta_k})^j \sum_{\ell = 0}^\infty \left(\frac{A}{\beta\e^{\im\theta_k}}\right)^\ell  \notag\\
    &= \sum_{j=0}^\infty\sum_{\ell=0}^\infty a_j\beta^{j - \ell} A^\ell \left(\frac{1}{M} \sum_{k=0}^{M-1}\e^{\frac{2\pi\im}{M}(j-\ell)k} \right) \notag\\
    &= \sum_{j=0}^\infty\sum_{\ell=0}^\infty a_j\beta^{j - \ell} A^\ell S_M(j-\ell),
  \end{align}
  where $S_M(y) = \frac{1}{M} \sum_{k=0}^{M-1}\e^{\frac{2\pi\im}{M}yk}$.
  Let $\Z_{\ge 0} = \{0, 1, 2, \dots \}$ be the set of all non-negative integers.
  Clearly, $\Z_{\ge 0}^2 =
  \{(j,\ell) \in \Z_{\ge 0}^2 \mid j < \ell \} \cup
  \{(j,\ell) \in \Z_{\ge 0}^2 \mid j = \ell \} \cup
  \{(j,\ell) \in \Z_{\ge 0}^2 \mid j > \ell \}
  $ holds.
  Considering this,
  approximation $f_M(A)$ can be described as follows:
  \begin{align}
    f_M(A)
    &=
    \sum_{j=0}^\infty\sum_{\ell = j+1}^\infty a_j\beta^{j-\ell} A^\ell S_M(j-\ell) +
    \sum_{j=0}^\infty a_jA^j +
    \sum_{\ell = 0}^\infty\sum_{j = \ell+1}^\infty a_j\beta^{j-\ell} A^\ell S_M(j-\ell)  \notag\\
    &=
    \sum_{j=0}^\infty\sum_{y = 1}^\infty a_j\beta^{-y} A^{j+y} S_M(-y) +
    f(A) +
    \sum_{\ell = 0}^\infty\sum_{y = 1}^\infty a_{\ell + y}\beta^y A^\ell S_M(y).
  \end{align}
  As $S_M(y)$ is equal to $1$ for the case in which $y$ is multiples of $M$, and $0$ otherwise, we have
  \begin{align}
    f_M(A) =
    \sum_{j=0}^\infty\sum_{y = 1}^\infty a_j\beta^{-yM} A^{j+yM} +
    f(A) +
    \sum_{\ell = 0}^\infty\sum_{y = 1}^\infty a_{\ell + yM}\beta^{yM}A^\ell.
  \end{align}
  Therefore, the error is bounded as
  \begin{align}
    \Norm{f(A) - f_M(A)}
    &=
    \Norm{
      \sum_{j=0}^\infty\sum_{y = 1}^\infty a_j\beta^{-yM} A^{j+yM} +
      \sum_{\ell = 0}^\infty\sum_{y = 1}^\infty a_{\ell + yM}\beta^{yM}A^\ell
    }  \notag\\
    &\le
    \sum_{j=0}^\infty\sum_{y = 1}^\infty \abs{a_j}\beta^{-yM} \norm{A}^{j+yM} +
    \sum_{\ell = 0}^\infty\sum_{y = 1}^\infty \abs{a_{\ell + yM}}\beta^{yM}\norm{A}^\ell.
  \end{align}
  From Cauchy's estimate $\abs{a_j} \le B/R^j$ and $\norm{A} < \beta < R$, we have
  \begin{align}
    \Norm{f(A) - f_M(A)}
    &\le
    \sum_{j=0}^\infty\sum_{y = 1}^\infty B\left(\frac{\norm{A}}{R}\right)^j  \left(\frac{\norm{A}}{\beta}\right)^{yM} +
    \sum_{\ell = 0}^\infty\sum_{y = 1}^\infty B\left(\frac{\norm{A}}{R}\right)^\ell \left(\frac{\beta}{R} \right)^{yM}  \notag\\
    &=
    \frac{B}{1 - \frac{\norm{A}}{R}}
    \left(
    \frac{1}{1 - \bigl(\frac{\norm{A}}{\beta} \bigr)^M }\left(\frac{\norm{A}}{\beta}\right)^M +
    \frac{1}{1 - \bigl(\frac{\beta}{R}        \bigr)^M}\left(\frac{\beta}{R}\right)^M
    \right).
  \end{align}
}

% ------------------------------------------------------------------------------
%\input{v36_QuantAlgo.tex}
\section{Quantum algorithm}\label{QuantAlgo}
In this section, we describe the quantum algorithm to compute state $\ket{f}$.
In Section \ref{sec:goal},
we explain that the goal of the quantum algorithm is to compute a weighted sum of
the solutions of linear systems.
In Section  \ref{sec:description}, we provide the description of the quantum algorithm.

\subsection{Goal of the algorithm}\label{sec:goal}
Our original goal is to output state
%===============================================================================
$\ket{f} = f(A)\ket{b}/\norm{f(A)\ket{b}}$.
%===============================================================================
To obtain state $\ket{f}$, we use approximation $f_M(A)$,
which approximates matrix function $f(A)$ with exponential accuracy;
that is, we consider constructing a quantum algorithm that outputs state
%===============================================================================
\begin{align}
  \ket{f_M} := \frac{f_M(A)\ket{b}}{\norm{f_M(A)\ket{b}}},
\end{align}
%===============================================================================
instead of state $\ket{f}$.
%===============================================================================
Matrix-vector product $f_M(A)\vec{b}$ can be described as
the sum of solutions $\vec{x}_k := (\e^{\im \theta_k}I_N - A/\beta)^{-1}\vec{b}$
with weight $g_k := f(\beta \e^{\im\theta_k})\e^{\im\theta_k}$, that is,
\begin{align}
    f_M(A)\vec{b}
    &= \frac{1}{M}\sum_{k=0}^{M-1} f(\beta\e^{\im \theta_k}) \beta\e^{\im \theta_k}( \beta\e^{\im \theta_k} I_N - A)^{-1}\vec{b}  \notag \\
    &= \frac{1}{M}\sum_{k=0}^{M-1} f(\beta\e^{\im \theta_k})\e^{\im \theta_k}( \e^{\im \theta_k} I_N - A/\beta)^{-1}\vec{b} \notag \\
    &=  \frac{1}{M}\sum_{k=0}^{M-1} g_k \vec{x}_k.
    \label{approx}
\end{align}
Thus, to obtain state $\ket{f_M}$,
we consider a quantum algorithm that computes weighted sum \eqref{approx} of the solutions
of the linear systems.

\subsection{Algorithm description}\label{sec:description}
We explain the procedure of the quantum algorithm that computes state $\ket{f_M}$.
To describe the algorithm, we do not consider the error in states.
We discuss the analysis of the error in Section \ref{errorOfAlgo}.

% ==============================================================================
% ==============================================================================
% ==============================================================================

\noindent
\hrulefill

\noindent
\textbf{Algorithm 1.}\\
\noindent
\hrulefill

% ==============================================================================
% ==============================================================================
% ==============================================================================
\noindent\textbf{Step 1.}
Apply the improved version \cite{CKS:2015} of the HHL algorithm to obtain quantum state
%===============================================================================
$\ket{x'}$
%===============================================================================
corresponding to solution $\vec{x'} \in \C^{NM}$ of block diagonal linear system
\begin{align}
  A'\vec{x}' = \vec{b}',
  \label{eq:Ax=bdash}
\end{align}
where $A' \in \C^{NM \times NM}$ is a block diagonal matrix defined as
\begin{align}
  A' :=
  \begin{bmatrix}
    \e^{\im\theta_0}I_N - A/\beta &                             &         & \\
                                & \e^{\im\theta_1}I_N - A/\beta &         & \\
                                &                             & \ddots  & \\
                                &                             &         & \e^{\im\theta_{M-1}}I_N - A/\beta \\
  \end{bmatrix},
  \label{Aprime}
\end{align}
and
\begin{align}
  \vec{x}' := \begin{bmatrix} \vec{x}_0 \\ \vec{x}_1 \\ \vdots \\ \vec{x}_{M-1} \end{bmatrix}, \qquad
  \vec{b}' := \begin{bmatrix} \vec{b} \\ \vec{b} \\ \vdots \\ \vec{b} \end{bmatrix} =
  \begin{bmatrix} 1 \\ 1 \\ \vdots \\ 1\end{bmatrix} \otimes \vec{b}.
  \label{defofxbprime}
\end{align}
Applying the HHL algorithm or its improved algorithm to the block diagonal linear system \eqref{eq:Ax=bdash},
we have state
\begin{align}
  \ket{x'}
  &
  = \qstate{\sum_i{x'}^{[i]}\ket{i}}
  = \sum_{k=0}^{M-1}\frac{ \norm{\vec{x}_k} }{ \norm{\vec{x'}}  }\ket{k}
   \left( \sum_{i=0}^{N-1}\frac{{x}_k^{[i]}}{\norm{\vec{x}_k}}\ket{i} \right)
  = \sum_{k=0}^{M-1}p_k\ket{k}\ket{x_k},
\end{align}
where $x'^{[i]}$ and $x_k^{[i]}$ is the $i$-th element of $\vec{x'}$ and $\vec{x}_k$, respectively, $p_k = \norm{\vec{x}_k}/\norm{\vec{x'}}$, and
%===============================================================================
$\ket{x_k}=\qstateinline{\sum_i {x}_k^{[i]}\ket{i}}$.

% ==============================================================================
% ==============================================================================
% ==============================================================================
\medskip
\noindent\textbf{Step 2.}
Add an ancilla qubit and perform unitary operator $U_f$ such that
\begin{align}
  U_f\ket{k}\ket{0} = \ket{k}\left(Cg_k\ket{0} + \sqrt{1 - \abs{Cg_k}^2}\ket{1} \right), \quad
  \label{eq:Uf}
\end{align}
for $k \in \{0,1,\dots,M-1\} $
to multiply weight $g_k = f(\beta\e^{\im\theta_k})\e^{\im\theta_k}$,
where $C$ is a constant such that $\abs{Cg_k} \le 1$.
(Actually, we use the unitary $U_{\tilde{f}_L}$ that is replaced $g_k$ with $\tilde{g}_k = \tilde{f}_L(\beta\e^{\im\theta_k})\e^{\im\theta_k}$,
where $\tilde{f}_L(z)$ is the truncated series of $f(z)$ at order $L$. )
Performing unitary operator $U_f$ to state $\ket{x'}\ket{0}$ yields quantum state
\begin{align}
  U_f \left(\sum_{k=0}^{M-1}p_k\ket{k}\ket{x_k}\ket{0}\right)
  = C\sum_{k=0}^{M-1}p_kg_k\ket{k}\ket{x_k} \ket{0} + \ket{\Phi_0^\perp},
\end{align}
where $\ket{\Phi_0^\perp}$ satisfies $(I_M \otimes I_N \otimes \ket{0}\bra{0})\ket{\Phi_0^\perp} = 0$.

% ==============================================================================
\medskip
\medskip
\noindent\textbf{Step 3.}
Apply the Hadamard gates $H^{\otimes m} \otimes I_N \otimes I_2$ to obtain the weighted sum.
We have
\begin{align}
  & \left( H^{\otimes m} \otimes I_N \otimes I_2 \right) \left( C\sum_{k=0}^{M-1}p_kg_k\ket{k}\ket{x_k} \ket{0} + \ket{\Phi_0^{\perp}} \right) \notag\\
  &= \frac{C}{\sqrt{M}}\sum_{k=0}^{M-1}p_kg_k\ket{0^m}\ket{x_k} \ket{0} + \ket{\Phi_0^{\prime \perp}} \notag\\
  &= \ket{0^m}\otimes \frac{C}{\sqrt{M}}\sum_{k=0}^{M-1}\frac{\norm{\vec{x}_k}}{\norm{\vec{x'}}}g_k\sum_{i=0}^{M-1}\frac{\vec{x}_k^{[i]}}{\norm{\vec{x}_k}}\ket{i}
  \otimes \ket{0} + \ket{\Phi_0^{\prime \perp}} \notag \\
  &= \ket{0^m}\otimes \frac{C\sqrt{M}}{\norm{\vec{x'}}}
  \sum_{i=0}^{N-1}  \left(\frac{1}{M}\sum_{k=0}^{M-1}g_k\vec{x}_k^{[i]}\right)
  \ket{i} \otimes \ket{0} + \ket{\Phi_0^{\prime \perp}} \notag \\
  &= \ket{0^m} \otimes \frac{C\sqrt{M}\norm{f_M(A)\vec{b}}}{\norm{\vec{x}'}} \ket{f_M} \otimes \ket{0} + \ket{\Phi_0^{\prime \perp}},
\end{align}
where $\ket{\Phi_0^{\prime \perp}}$ satisfies $(\gaiseki{0^m}{0^m} \otimes I_N \otimes \gaiseki{0}{0})\ket{\Phi_0^{\prime \perp}} = 0$.

% ==============================================================================
\medskip
\noindent\textbf{Step 4.}
Measure the first register and the ancilla qubit in the computational basis.
If we obtain an outcome $00\cdots 0$ of the first register and $0$ of the ancilla qubit,
then we have state $\ket{f_M} = \qstateinline{f_M(A)\ket{b}}$.

%==============================================================================

\noindent
\hrulefill
% ==============================================================================
% ==============================================================================
% ==============================================================================

As we can see from Eq.~\eqref{extendedlinearsystems},
we can obtain state $\ket{x'}$ even if $A$ is not Hermitian.
This is the reason that our algorithm can output state $\ket{f}$ even if $A$ is not Hermitian.

As we can see from Steps 1 and 2,
Algorithm 1, which is the proposed quantum algorithm, uses the HHL algorithm and unitary operator $U_f$ as subroutines.

To apply the HHL algorithm or its improved algorithm,
the norm of the coefficient matrix must be no more than $1$ from the problem setting (Definition \ref{QLSP}).
Additionally, we need to construct the oracle that returns elements of the coefficient matrix
and the oracle that returns the positions of the nonzero elements.
We also need an oracle to generate the state corresponding to the right-hand side vector of the linear system.
In Section \ref{sec:LS}, we consider this problem for the block diagonal linear system \eqref{eq:Ax=bdash}.

Subroutine $U_f$ is defined by the infinite series $f(z) = \sum_{j=0}^\infty a_jz^j$.
To manage the infinite series, we consider truncated series $\tilde{f}_L(z) := \sum_{j=0}^{L-1}a_jz^j$;
that is, instead of unitary operator $U_f$,
we consider unitary operator $U_{\tilde{f}_L}$ that is replaced
weight $g_k = f(\beta\e^{\im\theta_k})\e^{\im\theta_k}$ in the unitary operator $U_f$ with $\tilde{g}_k = \tilde{f}_L(\beta\e^{\im\theta_k})\e^{\im\theta_k}$.
In Section \ref{sec:Uf}, we show the procedure and runtime of the unitary operator $U_{\tilde{f}_L}$.

To describe the algorithm, we did not consider the error.
However, the actual output state of the HHL algorithm includes the error.
Additionally, we need to consider the approximation error of state $\ket{f_M}$ and unitary operator $U_{\tilde{f}_L}$.
In Section \ref{ErrorRuntime}, we provide an error analysis of the algorithm, that is,
we provide an upper bound of $\norm{\ket{f} - \ket{\tilde{f}}}$,
where $\ket{\tilde{f}}$ is the quantum state that is actually outputted.
Moreover, we show a runtime and lower bound of the success probability, and provide the proof of the main theorem (Theorem \ref{MainResult}).

% ------------------------------------------------------------------------------
% \input{v36_LS.tex}
\section{Linear system $A'\vec{x'} = \vec{b'}$ in Step 1 of the quantum algorithm}\label{sec:LS}
In this section, we discuss applying the HHL algorithm
to linear system $A'\vec{x'} = \vec{b'}$ in Step 1 of Algorithm 1 in detail,
and derive the following proposition.

\begin{proposition}\label{prop:LS}
  For quantum state $\ket{x'} = \qstateinline{ \sum_i {x'}^{ [i]}\ket{i} }$
  corresponding
  to solution $\vec{x'}$ of linear system $A'\vec{x'} = \vec{b'}$ in Eq.~\eqref{eq:Ax=bdash}
  and some positive constant $\epsilon'$ such that $0 \le \epsilon' \le 1/2$,
  state $\ket{\tilde{x}'}$ such that $\norm{\ket{x'} - \ket{\tilde{x}'}}\le \epsilon'$ can be obtained using
  \begin{align}
    O\left(d\kappa'^2 \log^2\left(\frac{d\kappa'}{\epsilon'}\right)\right)
    \text{ queries to oracle $\mathcal{P}_A$ and }
    O\left(\kappa' \log\left(\frac{d\kappa'}{\epsilon'} \right) \right)
    \text{ uses of $\mathcal{P}_{\vec{b}}$},
    \label{eq:lemmaLS}
  \end{align}
  with gate complexity
  \begin{align}
    O\left(d\kappa'^2 \log^2\left(\frac{d\kappa'}{\epsilon'}\right)
    \left[\log(NM) +
    \log^{\frac{5}{2}}\left(\frac{d \kappa'}{\epsilon'}\right)\right] \ \right),
    \label{eq:lemmaLSgate}
  \end{align}
  where $d$ is the sparsity of matrix $A$ and $\kappa' := 1/(1 - \beta^{-1})$.
\end{proposition}
This proposition is used in the proof of Corollary \ref{Runtime}, which shows the complexity of Algorithm 1.

The remainder of this section is as follows:
In Section \ref{sec:norms}, we derive the upper bounds of $\norm{A'}$, $\norm{A'^{-1}}$
and condition number $\kappa_{A'} := \norm{A'}\norm{A'^{-1}}$ of matrix $A'$.
In Section \ref{sec:oracle},
we explain that oracle $\mathcal{P}_{A'}$ for matrix $A'$ can be constructed using $\O(1)$ queries to $\mathcal{P}_A$.
Additionally, we explain the gate complexity.
Section \ref{sec:proofOfLemma} is devoted to the proof of Proposition \ref{prop:LS}.

\subsection{Upper bounds of $\norm{A'}, \norm{A^{\prime -1}}$ and the condition number of $A'$}\label{sec:norms}
First, we present the upper bound of $\norm{A'}$.
Note that $\norm{X}$ represents the spectral norm of square matrix $X$.
Thus, $\norm{X}$ is equal to the largest singular value of square matrix $X$.
\begin{lem}\label{normOfAprime}
  Let $\norm{A} \le 1 < \beta$.
  Then, for matrix $A'$ in Eq.~\eqref{Aprime}, $\norm{A'} \le 1 + \beta^{-1} < 2$ holds.
\end{lem}
\myproof{
  Matrix $A'$ can be represented as $A' = \diag(\e^{\im\theta_0},\e^{\im\theta_1},\dots,\e^{\im\theta_{M-1}})\otimes I_N - I_M \otimes A/\beta$.
  For square matrices $X$ and $Y$, $\norm{X \otimes Y} = \norm{X}\norm{Y}$ holds.
  Thus,
  \begin{align}
    \norm{A'}
    &=
    \norm{\diag(\e^{\im\theta_0},\e^{\im\theta_1},\dots,\e^{\im\theta_{M-1}})\otimes I_N - I_M \otimes A/\beta} \notag \\
    &\le
    \norm{\diag(\e^{\im\theta_0},\e^{\im\theta_1},\dots,\e^{\im\theta_{M-1}})\otimes I_N} +
    \norm{I_M \otimes A/\beta} \notag \\
    & \le 1 + \beta^{-1} \notag \\
    & < 2,
  \end{align}
  where the second inequality uses $\norm{A} \le 1$.
}

Next, we show the upper bound of $\norm{A'^{-1}}$.
\begin{lem}\label{normOfAprimeInv}
  Let $\norm{A} \le 1 < \beta$.
  Then, for matrix $A'$ in Eq.~\eqref{Aprime}, $\norm{A^{\prime -1}} \le (1 - \beta^{-1})^{-1}$ holds.
\end{lem}
\proof{
  Because $A'$ is a block diagonal matrix, $A^{\prime -1}$ is also a block diagonal matrix
  that the diagonal blocks are $(\e^{\im\theta_k}I_N - A/\beta)^{-1} \ (k \in \{0,1,\dots,M-1\})$.
  Thus, we have $\norm{A^{\prime -1}} = \max\{ \norm{(\e^{\im\theta_k}I_N - A/\beta)^{-1}} \mid k \in \{0,1,\dots,M-1\} \}$.
  As is well known,
  \begin{align}
    \norm{(\e^{\im\theta_k}I - A/\beta)^{-1}} \le (1 - \norm{A}/\beta)^{-1}
  \end{align}
  holds. Hence,
  \begin{align}
    \norm{A^{\prime -1}}
    &= \max\left\{ \norm{(\e^{\im\theta_k}I - A/\beta)^{-1}} \mid k \in \{0,1,\dots,M-1\} \right\} \notag \\
    &\le (1 - \norm{A}/\beta)^{-1} \notag \\
    &\le (1- \beta^{-1})^{-1},
  \end{align}
  where the second inequality uses $\norm{A} \le 1$.
}

This lemma is also used in the proof of Propositions \ref{Manddelta} and \ref{prop:successprob}
to show the error and success probability of Algorithm 1.

Finally, we describe the upper bound of condition number $\kappa' = \norm{A^{\prime}}\norm{A^{\prime -1}}$ of matrix $A'$.
This bound is obtained immediately from Lemmas \ref{normOfAprime} and \ref{normOfAprimeInv}.

\begin{cor}\label{cor:conditionNumberOfAprime}
  Let $\norm{A} \le 1 < \beta$.
  Condition number $\kappa_{A'} := \norm{A'}\norm{A^{\prime -1}}$ of block diagonal matrix $A'$ in Eq.~\eqref{Aprime}
  is bounded as $\kappa_{A'} < 2\kappa'$, where $\kappa' = (1 - \beta^{-1})^{-1}$.
\end{cor}

\subsection{Oracle that accesses matrix $A'$}\label{sec:oracle}
To apply the HHL algorithm to block diagonal linear system $A'\vec{x'} = \vec{b'}$,
oracle $\mathcal{P}_{A'}$ is required to access the elements of coefficient matrix $A'$.
Oracle $\mathcal{P}_{A'}$ consists of oracle $O_{\nu'}$ that returns the position of the nonzero element of $A'$
and oracle $O_{A'}$ that returns the element of $A'$ for a given position.
In this section, we consider the construction of $\mathcal{P}_{A'}$ under the assumption that
we can use oracle $\mathcal{P}_A$.
Furthermore, we show that oracle $\mathcal{P}_{A'}$ can be performed using $\O(1)$ queries to oracle $\mathcal{P}_{A}$.
%===============================================================================
Additionally, we consider the gate complexity of $O_{\nu'}$ and $O_{A'}$.
Here, for simplicity, we assume that the all diagonal elements of $A$ are nonzero.
%===============================================================================

Let us consider oracle $O_{\nu'}$ that returns the position of the nonzero element of $A'$.
Because $A'$ is the block diagonal matrix defined by Eq.~\eqref{Aprime},
%===============================================================================
%===============================================================================
the row index of the $\ell$-th nonzero element in the $(kN+j)$-th column is $kN + \nu(j,\ell)$
for $k \in \{0,1,\dots,M-1\}$ and $j \in \{0,1,\dots,N-1\}$.
Thus, $O_{\nu'}$ can be constructed using oracle $O_{\nu}$ and CNOT gates:
$\ket{k,j}\ket{0^m, \ell}
\mapsto_{O_{\nu}} \ket{k,j}\ket{0^m, \nu(j,\ell)}
\mapsto \ket{k,j}\ket{k,\nu(j,\ell)} = O_{\nu'}\ket{k,j}\ket{0^m, \ell}$.
Therefore, the query complexity of $O_{\nu'}$ is $O(1)$ and the gate complexity is $O(m) = O(\log(M))$.
%===============================================================================

%===============================================================================
Next, we consider constructing the unitary operator that
returns the element of $A'$ for a given position.
Again, focusing on that $A'$ is the block diagonal matrix,
we can see that the elements in the non-diagonal blocks are always zero.
Thus, it is sufficient to consider the construction of unitary operator $O_{A'}$ such that
%===============================================================================
$O_{A'}\ket{k,i}\ket{k',j}\ket{0} = \ket{k,i}\ket{k',j}\ket{0}$ for $k \neq k'$ and
%===============================================================================
\begin{align}
  O_{A'}\ket{k,i}\ket{k,j}\ket{0} &= \ket{k,i}\ket{k,j}\ket{  \delta_{i,j}\e^{\im\theta_k} - A_{i,j}/\beta }, %\\
  \label{oracleA}
\end{align}
where
%===============================================================================
the value written inside $\ket{ \delta_{i,j}\e^{\im\theta_k} - A_{i,j}/\beta }$
means the binary representation of $\delta_{i,j}\e^{\im\theta_k} - A_{i,j}/\beta$
and
%===============================================================================
$\delta_{i,j}$ denotes the Kronecker delta,
that is, $\delta_{i,i} = 1$ and $\delta_{i,j} \neq 0$ for $i \neq j$.
As we can see from Eq.~\eqref{oracleA},
oracle $O_{A'}$ is constructed using $\O(1)$ uses of
oracle $O_A$, the adder, multiplier,
comparator
$\ket{k,k'}\ket{0} \mapsto \ket{k,k'}\ket{\delta_{k,k'}}, \ket{i,j}\ket{0} \mapsto \ket{i,j}\ket{\delta_{i,j}}$,
and quantum arithmetic circuit to obtain quantum state $\ket{\e^{\im\theta_k}}$,
where the value written inside $\ket{\e^{\im\theta_k}}$ means the binary representation of $\e^{\im\theta_k}$.
%===============================================================================
%===============================================================================
Specifically, $O_{A'}$ can be performed by the following procedure.
Here, we use quantum registers
$\ket{0}_{\textrm{r1}}, \ket{0}_{\textrm{r2}}, \ket{0}_{\textrm{r3}}, \ket{0}_{\textrm{r4}}, \ket{0}_{\textrm{r5}}$
and flag qubits $\ket{0}_{\textrm{f1}}, \ket{0}_{\textrm{f2}}$.
% All internal calculation is performed exactly.
\begin{enumerate}
  \item
  First, perform comparator $\ket{k,k'}\ket{0}_{\textrm{f1}} \mapsto \ket{k,k'}\ket{\delta_{k,k'}}_{\textrm{f1}}$
  to check whether the values of $k$ and $k'$ are the same.
  Next, conditioned on the `f1' qubit being $1$, perform comparator $\ket{i,j}\ket{0}_{\textrm{f2}} \mapsto  \ket{i,j}\ket{\delta_{i,j}}_{\textrm{f2}}$ to check whether the values of $i$ and $j$ are the same.
  \item
  Conditioned on the `f1' qubit being $1$,
  perform
  $\ket{i,j}\ket{0}_{\textrm{r1}}\ket{0}_{\textrm{r2}}
  \mapsto \ket{i,j}\ket{A_{i,j}}_{\textrm{r1}}\ket{0}_{\textrm{r2}}
  \mapsto \ket{i,j}\ket{A_{i,j}}_{\textrm{r1}}\ket{-A_{i,j}/\beta}_{\textrm{r2}}$
  using oracle $O_A$ and the multiplier to obtain $-A_{i,j}/\beta$.
  \item
  Conditioned on the `f1' and `f2' qubit being $1$,
  perform
  $
  \ket{k}\ket{0}_{\textrm{r3}}\ket{0}_{\textrm{r4}}
  \mapsto \ket{k}\ket{\theta_k= \frac{2\pi}{M}k}_{\textrm{r3}}\ket{0}_{\textrm{r4}}
  \mapsto \ket{k}\ket{\theta_k}_{\textrm{r3}}\ket{\e^{\im\theta_k}}_{\textrm{r4}}
  $
  using the multiplier and the quantum arithmetic circuit to compute $\e^{\im\theta_k}$.
  \item
  Conditioned on the `f1' qubit being $1$,
  add the values of `r2' and `r4' register and write the result into the `r5' register.
  When the value of `f1' qubit is $1$, the value stored in the `r2' register is $-A_{i,j}/\beta$.
  Therefore, then, $\ket{-A_{i,j}/\beta  }_{\textrm{r2}}
   \ket{v}_{\textrm{r4}}
   \ket{0}_{\textrm{r5}}
   \mapsto
   \ket{-A_{i,j}/\beta  }_{\textrm{r2}}
    \ket{v}
    \ket{v -A_{i,j}/\beta  }_{\textrm{r5}}$
  is performed using the adder,
  where
  $v$ represents the value stored in the `r4' register,
  that is, $v = \e^{\im\theta_k}$ when the value of the `f2' qubit is $1$
  and $v=0$ when the  value of the `f2' qubit is $0$.
  \item
  We uncompute the `r1',`r2',`r3',`r4' registers and the `f1',`f2' qubits.
  Then, the value stored in the `r5' register represents $0$ for $k \neq k'$ and
  $\delta_{i,j}\e^{\im\theta_k} - A_{i,j}/\beta$ for $k = k'$.
\end{enumerate}
%===============================================================================
We consider the gate complexity when $O_{A'}$ returns the element of $A'$ with $s$ bits of accuracy.
In Step 1, the gate complexity of comparator
$\ket{k,k'}\ket{0} \mapsto \ket{k,k'}\ket{\delta_{k,k'}}$ and
$\ket{i,j}\ket{0} \mapsto \ket{i,j}\ket{\delta_{i,j}}$ are $O(\log (M))$ and $O(\log(N))$, respectively.
In Step 2, if oracle $O_{A}$ outputs the element with $s$ bits of accuracy and constant $1/\beta$ has $s$ bits,
then we can obtain the value of $-A_{i,j}/\beta$ with $s$ bits of accuracy using $O(s^2)$ gates.
In Step 3, by multiplying $2k/M$ by $\pi$ with $s$ bits of accuracy,
we can obtain $\theta_k$ with $s$ bits of accuracy using $O(s^2)$ gates.
% Therefore, $\e^{\im\theta_k}$ is calculated with $s$ bits of accuracy.
Therefore, the binary representation of $\e^{\im\theta_k}$, which is the output of the quantum arithmetic circuit, is calculated with $s$ bits of accuracy.
Using techniques based on Taylor series and long multiplication, this calculation can be performed with $O(s^{\frac{5}{2}})$.
In Step 4, we use $O(s)$ gates to obtain $\ket{\e^{\im\theta_k}-A_{i,j}/\beta}$ with $s$ bits of accuracy.
Thus, the overall gate complexity of $O_{A'}$ to output the element of $A'$ with $s$ bits of accuracy is
$O(\log(MN)  + s^{\frac{5}{2}})$.
%===============================================================================
We summarize this in the following lemma.
\begin{lem}\label{remark:PAprime}
  Oracle $\mathcal{P}_{A'}$ can be performed using $\O(1)$ queries to oracle $\mathcal{P}_A$.
  Moreover,
  the gate complexity of $O_{\nu'}$ is $O(\log(M))$
  and the gate complexity of $O_{A'}$ to output the element of $A'$ with $s$ bits of accuracy is $O( \log(NM) + s^{\frac{5}{2}} )$.
\end{lem}

\subsection{Applying the HHL algorithm to $A'\vec{x'} = \vec{b'}$ }\label{sec:proofOfLemma}
We provide the proof of Proposition \ref{prop:LS}.
To apply the HHL algorithm to linear systems,
the norm of the coefficient must be no more than $1$, from the problem setting.
Additionally, we need the oracles for the coefficient matrix and the right-hand side vector.

The construction of oracle $\mathcal{P}_{A'}$ for matrix $A'$ has been discussed (Lemma \ref{remark:PAprime}).
In the following, we consider the problem for the norm of block diagonal matrix $A'$
and oracle $\mathcal{P}_{\vec{b'}}$ for right-hand side vector $\vec{b'}$.

\noindent\textbf{Proof of Proposition \ref{prop:LS}.}
First, to bound the norm of the coefficient, we consider scaling.
We rewrite linear system $A'\vec{x'} = \vec{b'}$ as linear system $(A'/c)(c\vec{x'}) = \vec{b'}$
using particular  constant $c$ (e.g., $c=2$).
Clearly, the state corresponding to the solution of the rewritten linear system is state $\ket{x'}$.
Thus, we consider applying the HHL algorithm to the rewritten linear system, $(A'/c)(c\vec{x'}) = \vec{b'}$, instead of the original
linear system, $A'\vec{x'} = \vec{b'}$.
Considering the multiplier, the queries to  oracle $\mathcal{P}_{A'/c}$ are equal
to the queries to oracle $\mathcal{P}_{A'}$.
Because $\norm{A'} \le 1 + \beta^{-1}$ from Lemma \ref{normOfAprime},
$\norm{(A'/c)} \le 1$ holds if constant $c$ is chosen such that $c \ge 1 + \beta^{-1}$ (e.g., $c=2$).

Next, we consider oracle $\mathcal{P}_{\vec{b'}}$
that generates state $\ket{b'} = \qstateinline{\sum_i {b'}^{[i]}\ket{i}}$,
that is, we consider oracle $\mathcal{P}_{\vec{b'}}$ such that $\mathcal{P}_{\vec{b'}}\ket{0^{nm}} = \ket{b'}$,
where $b'^{[i]}$ is the $i$-th element of $\vec{b'}$.
%=-------------------------------------
From the definition (Eq.~\eqref{defofxbprime}) of vector $\vec{b'}$, oracle $\mathcal{P}_{\vec{b'}}$ can be described as
\begin{align}
   \mathcal{P}_{\vec{b'}} = H^{\otimes m}\otimes \mathcal{P}_{\vec{b}}.
\end{align}
Thus, the number of queries to $\mathcal{P}_{\vec{b'}}$ is equal to the number of queries to oracle $\mathcal{P}_{\vec{b}}$.
Furthermore, the gate complexity is $O(\log(NM))$.
From the above and Lemma \ref{remark:PAprime}, the conditions for applying the HHL algorithm to $(A'/c)(c\vec{x'}) = \vec{b'}$ are satisfied.

%===============================================================================
We consider the query complexity.
From Lemma \ref{remark:PAprime} and the above,
oracle $\mathcal{P}_{A'/c}$ and $\mathcal{P}_{\vec{b'}}$ can be performed using $\O(1)$ queries to oracle $\mathcal{P}_A$ and $\mathcal{P}_{\vec{b}}$.
Furthermore, the upper bound of condition number $\kappa_{A'}$
of matrix $A'$ is obtained from Corollary \ref{cor:conditionNumberOfAprime}.
This implies the query complexity \eqref{eq:lemmaLS}.

We explain the gate complexity.
In the improved version of the HHL algorithm \cite[Theorem 4]{CKS:2015},
the oracle for the matrix is used to perform a walk operator
(for the details of the definition of walk operator, see \cite{CKS:2015,BCK:2015}).
From the proof of \cite[Lemma 10]{BCK:2015},
the gate complexity of the walk operator for $A'/c$ is the sum of $O(\log (NM))$,
the gate complexity of $O_{A'/c}, O_{\nu'}$,
and the gate complexity of computing square root and trigonometric functions based on the output of $O_{A'/c}$.
The gate complexity oracle $O_{A'/c}$ are equal to the gate complexity of oracle $O_{A'}$.
Moreover, when oracle $O_{A'}$ outputs the element with $s$ bits of accuracy,
oracle $O_{A'/c}$ also outputs the elements with $s$ bits of accuracy.
Thus, the gate complexity of computing square root and trigonometric functions is $O( s^{\frac{5}{2}} )$
using techniques based on Taylor series and long multiplication \cite{BCK:2015}.
Therefore, from Lemma \ref{remark:PAprime},
the gate complexity of the walk operator for $A'/c$ is $O(\log(NM) + s^\frac{5}{2})$
when $O_{A'}$ outputs the element of $A'$ with $s$ bits of accuracy.
Hence, as shown in the proof of \cite[Theorem 4]{CKS:2015},
a step of the quantum walk for $A'$ can be performed within error $\epsilon''$ with gate complexity
$O(\log(NM)  + \log^\frac{5}{2}\left(d\kappa_{A'}/\epsilon''\right))$.
%===============================================================================
Thus, the proposition holds from the proof of Proposition \ref{ImprovedVersionOfHHL}(\cite[Theorem 4]{CKS:2015})
on the improved version of the HHL algorithm.
\quad \qed

% ------------------------------------------------------------------------------
% \input{v36_Weight.tex}
\section{Unitary operator for multiplying the weight in Step 2 of the quantum algorithm}\label{sec:Uf}
In this section, we describe the construction of
unitary operator $U_{\tilde{f}_L}$ that replaces weight $g_k$ of unitary operator $U_f$
with $\tilde{g}_k$, and derive the following proposition,
where $\tilde{g}_k = \tilde{f}_L(\beta\e^{\im\theta_k})\e^{\im\theta_k}$ and
$\tilde{f}_L(z)$ is the truncated series of function $f(z)$ at order $L$.

\begin{proposition}\label{prop:Uf}
For positive integers $m$ and $\ell$, let $M = 2^m$ and $L = 2^\ell$, respectively.
Let $V$ be a unitary operator such that $V\ket{k}\ket{j} = (\e^{\im\theta_k})^{j+1}\ket{k}\ket{j}$
for quantum registers $\ket{k}, \ket{j} \ (k \in \{0,1,\dots,M-1\}, \ j \in\{0,1,\dots,L-1\})$,
where $\theta_k = 2\pi k/M$.
For truncated series $\tilde{f}_L(z) = \sum_{j=0}^{L-1} a_jz^j $
of complex function $f(z)$,
let $\tilde{g}_k = \tilde{f}_L(\beta\e^{\im\theta_k})\e^{\im\theta_k}$ and
let $\alpha = \sum_{j=0}^{L-1}|a_j|\beta^j$. %\abs{a_0} + \abs{a_1}\beta + \cdots + \abs{a_{L-1}}\beta^{L-1}$.
Let $W$ and $W'$ be unitary operators such that
$W\ket{0^\ell} = \frac{1}{\sqrt{\alpha}}\sum_{j=0}^{L-1}\sqrt{a_j\beta^j}\ket{j}$ and
$W'\ket{0^\ell} = \frac{1}{\sqrt{\alpha}}\sum_{j=0}^{L-1}\sqrt{a_j^\ast \beta^j}\ket{j}$, respectively,
where $z^\ast$ denotes the complex conjugate of $z \in \C$.
We define $U_{\tilde{f}_L} := (I_{M} \otimes W^{\prime{\dagger}})V(I_{M} \otimes W)$.
Then,
\begin{align}
  U_{\tilde{f}_L}\ket{k}\ket{0^\ell}
  = \ket{k}\otimes \left(\tilde{C}\tilde{g}_k\ket{0^\ell} + \sqrt{1 - |\tilde{C}\tilde{g}_k|^2}\ket{\Psi_0^\perp}\right),
\end{align}
where $\tilde{C} = 1/\alpha$ and $\ket{\Psi_0^\perp}$ satisfies $\gaiseki{0^\ell}{0^\ell}\ket{\Psi_0^\perp} = 0$.
Additionally, the gate complexity of $U_{\tilde{f}_L}$ is
%=========================================================================
$O(L + \log(M)\log(L))$.
%=========================================================================
\end{proposition}

This proposition is based on the technique for the linear combination of unitaries (see e.g., \cite{Kot:2014}).
This proposition is used to show Corollary \ref{Runtime}, which describes the complexity of Algorithm 1.
Before starting the proof of the proposition, we present the following lemma.

\begin{lem}\label{lemma:phase}
  For positive integers $m$ and $\ell$, let $M=2^m$ and $L=2^\ell$, respectively.
  Let $V$ be a unitary operator such that $V\ket{k}\ket{j} = (\e^{\im\theta_k})^{j+1}\ket{k}\ket{j}$
  for quantum registers $\ket{k}, \ket{j} \ (k \in \{0,1,\dots,M-1\}, \ j \in\{0,1,\dots,L-1\})$,
  where $\theta_k = 2\pi k/M$.
  Then, the gate complexity of unitary operator $V$ is $O(\log(M)\log(L))$.
\end{lem}
\proof{
  We represent $\ket{k}$ and $\ket{j}$ as $\ket{k} = \ket{k_{m-1}}\ket{k_{m-2}}\cdots\ket{k_0}$
  and $\ket{j} = \ket{j_{\ell-1}}\ket{j_{\ell-2}}\cdots\ket{j_0}$, respectively,
  where $k_s,j_t \in \{0,1\}, s\in\{0,1,\dots,m-1\}$, and $t\in\{0,1,\dots,\ell-1\}$.
  Then, $k$ and $j$ can be represented as $k = \sum_{s = 0}^{m-1} k_s2^s$ and $j = \sum_{t = 0}^{\ell - 1}j_t 2^t$, respectively.
  We have $(\e^{\im\theta_k})^j = \exp(\im 2\pi kj / M)$ because $\theta_k = 2\pi k/M$.
  Thus,
  \begin{align}
    \exp\left(\im\frac{2\pi}{M} kj\right)
    = \exp\left(\sum_{s=0}^{m-1}\sum_{t=0}^{\ell - 1} \im \frac{2\pi}{M} k_sj_t 2^{s+t} \right)
    = \prod_{s=0}^{m-1}\prod_{t=0}^{\ell -1} \exp\left(\im \frac{2\pi}{M} k_sj_t 2^{s+t} \right).
  \end{align}
  Therefore, map $\ket{k}\ket{j}\mapsto (\e^{\im\theta_k})^j \ket{k}\ket{j}$ can be performed using
  $\O(m\ell) = \O(\log(M)\log(L))$ controlled gates.
  Additionally, for any $\ket{j}$, map $\ket{k}\ket{j} \mapsto \e^{\im\theta_k}\ket{k}\ket{j}$ can be performed
  with $O(\log M)$ gates. Thus, the lemma follows.
}

Now, we provide the proof of Proposition \ref{prop:Uf}.

\noindent
\textbf{Proof of Proposition \ref{prop:Uf}. }
  As shown in \cite{Kot:2014}, we show the proposition by straightforward computation.
  \begin{align}
    U_{\tilde{f}_L}\ket{k}\ket{0^\ell} =
    (I_M \otimes W^{\prime{\dagger}})V(I_M \otimes W)\ket{k}\ket{0^\ell}
    &= (I_M \otimes W^{\prime {\dagger} }) \frac{1}{\sqrt{\alpha}} \sum_{j=0}^{L-1}\sqrt{ a_j \beta^j}  V \ket{k}\ket{j} \notag\\
    &= (I_M \otimes W^{\prime {\dagger} }) \frac{1}{\sqrt{\alpha}} \sum_{j=0}^{L-1}\sqrt{ a_j \beta^j} (\e^{\im\theta_k})^{j+1} \ket{k}\ket{j} \notag\\
    &= \ket{k} \otimes \left(  W^{\prime {\dagger} } \frac{\e^{\im\theta_k}}{\sqrt{\alpha}} \sum_{j=0}^{L-1} \sqrt{ a_j \beta^j} (\e^{\im\theta_k})^j \ket{j} \right).
  \end{align}
  Unitary operator $ W^{\prime {\dagger} }$ can be described as follows:
  \begin{align}
    W^{\prime {\dagger} }
    = \gaiseki{0^\ell}{0^\ell} W^{\prime {\dagger} } + \left(I_L - \gaiseki{0^\ell}{0^\ell} \right)  W^{\prime {\dagger} }
    = \ket{0^\ell} \frac{1}{\sqrt{\alpha}}\sum_{j=0}^{L-1}\left( \sqrt{a_j^\ast \beta^j}\right)^\ast \bra{j} + (I_L - \gaiseki{0^\ell}{0^\ell}) W^{\prime {\dagger} }.
  \end{align}
  Note that $\beta > 0$ and $(\sqrt{a_j^\ast})^\ast = \sqrt{a_j}$.
  Thus, we have
  \begin{align}
    U_{\tilde{f}_L}\ket{k}\ket{0^\ell}
    &= \ket{k}\otimes \left(\frac{\e^{\im\theta_k}}{\alpha}\sum_{j=0}^{L-1}a_j(\beta\e^{\im\theta_k})^j \ket{0^\ell}
       + (I_L - \gaiseki{0^\ell}{0^\ell}) W^{\prime {\dagger} }\frac{\e^{\im\theta_k}}{\sqrt{\alpha}} \sum_{j=0}^{L-1} \sqrt{ a_j \beta^j} (\e^{\im\theta_k})^j \ket{j}      \right) \notag\\
    &= \ket{k}\otimes \left(\tilde{C}\tilde{g}_k\ket{0^\ell} + \sqrt{1 - |\tilde{C}\tilde{g}_k|^2}\ket{\Psi_0^\perp}\right),
  \end{align}
  where $\tilde{C} = 1/\alpha$ and $\ket{\Psi_0^\perp}$ satisfies $\gaiseki{0^\ell}{0^\ell}\ket{\Psi_0^\perp} = 0$.

  We consider the complexity of unitary operator $U_{\tilde{f}_L}$.
  Unitary operators $W, W'$ can be performed using $O(L)$ gates \cite{SBM:2006}.
  The gate complexity of unitary operator $V$ is $O(\log(M)\log(L))$ from Lemma \ref{lemma:phase}.
  Thus, the proposition follows.
\qed

% ------------------------------------------------------------------------------
% \input{v36_ErrorRuntime.tex}
\section{Analysis of the quantum algorithm and proof of the main result}\label{ErrorRuntime}
In this section, we analyze the complexity, error
and success probability of the proposed quantum algorithm (Algorithm 1).
In Section \ref{subsec:Runtime}, we describe the complexity.
In Section \ref{errorOfAlgo}, we analyze the error
and derive the setting of parameters to bound the error using positive constant $\epsilon$.
In Section \ref{subsec:SuccessProb}, we describe the success probability.
Finally, we provide the proof of our main theorem (Theorem \ref{MainResult}).

\subsection{Runtime}\label{subsec:Runtime}
Algorithm 1 consists of Steps 1 to 4.
Because Step 4 is the measurement process,
the complexity of Algorithm 1 becomes the complexity up to Step 3.
Therefore, we have the following corollary using
Proposition \ref{prop:LS} for solving linear systems in Step 1 and
Proposition \ref{prop:Uf} for multiplying the weights in Step 2.

\begin{cor}[Runtime of Algorithm 1]\label{Runtime}
  Consider the quantum algorithm described in Section \ref{QuantAlgo}.
  We assume that the improved version of the HHL algorithm in Step 1 outputs quantum state $\ket{\tilde{x}'}$
  such that $\norm{\ket{x'} - \ket{\tilde{x}'}}\le \epsilon'$, where $\epsilon'$ is the positive constant.
  We assume that we used $U_{\tilde{f}_L}$ instead of $U_f$ in Step 2.
  Then, to implement Algorithm 1, we need
  \begin{align}
    O\left(d\kappa'^2 \log^2\left(\frac{d\kappa'}{\epsilon'}\right)\right)
    \text{ queries to oracle $\mathcal{P}_A$ and }
    O\left(\kappa' \log\left(\frac{d\kappa'}{\epsilon'} \right) \right)
    \text{ uses of $\mathcal{P}_{\vec{b}}$}.
  \label{queries}
  \end{align}
  Furthermore, the gate complexity of Algorithm 1 is
 \begin{align}
   O\left(
     d\kappa'^2 \log^2
     \left( \frac{d\kappa'}{\epsilon'} \right)
     \left[ \log(NM)  +
     \log^{\frac{5}{2}}\left(\frac{d \kappa'}{\epsilon'}\right) \right]
     +
     L + \log(M)\log(L)
   \right),
   \label{gates}
 \end{align}
where $d$ is the sparsity of $A$ and $\kappa' = 1/(1 - \beta^{-1})$.
\end{cor}
\proof{
The query complexity and gate complexity of Step 1 are Eqs.~\eqref{eq:lemmaLS}
and \eqref{eq:lemmaLSgate}, respectively, in Proposition \ref{prop:LS}.
The gate complexity of Step 2 is $O(L + \log(M)\log(L))$ from  Proposition \ref{prop:Uf}.
In Step 3, we only apply $m = \log M$ Hadamard gates to the quantum register.
Thus, the proposition follows.
}

\subsection{Error}\label{errorOfAlgo}
We analyze the error of Algorithm 1.
Specifically, we derive the upper bound
of the distance between states $\ket{f}$ and $\ket{\tilde{f}}$,
where $\ket{\tilde{f}}$ is the actual output state of Algorithm 1 and approximates $\ket{f_M}$.
To derive the upper bound, we use the following lemma.
\begin{lem}\label{lemma:veclemma}
  For any vectors $\vec{v} \in \C^N$ and $\vec{w} \in \C^N$,
  the following inequality holds.
  \begin{align}
      \Norm{\frac{\vec{v}}{\Norm{\vec{v}}} - \frac{\vec{w}}{\Norm{\vec{w}}}} \le 2\frac{\Norm{\vec{v} - \vec{w}}}{\Norm{\vec{v}}}
      \label{keta-ketb}.
  \end{align}
\end{lem}
\proof{
  This proof is based on the proof of Proposition 9 in \cite{CKS:2015}.
  Using the triangle inequality,
  \begin{align}
      \Norm{\frac{\vec{v}}{\Norm{\vec{v}}} - \frac{\vec{w}}{\Norm{\vec{w}}}}
      & = \Norm{\frac{\vec{v}}{\Norm{\vec{v}}} -\frac{\vec{w}}{\Norm{\vec{v}}} + \frac{\vec{w}}{\Norm{\vec{v}}}- \frac{\vec{w}}{\Norm{\vec{w}}}} \notag \\
      & \le \Norm{\frac{\vec{v}}{\Norm{\vec{v}}} -\frac{\vec{w}}{\Norm{\vec{v}}} } +\Norm{ \frac{\vec{w}}{\Norm{\vec{v}}}- \frac{\vec{w}}{\Norm{\vec{w}}}} \notag \\
      & \le \frac{\Norm{\vec{v} - \vec{w}} }{\Norm{\vec{v}}} + \abs{\frac{1}{\Norm{\vec{v}}} - \frac{1}{\Norm{\vec{w}}}}\Norm{\vec{w}} \notag \\
      & =   \frac{\Norm{\vec{v} - \vec{w}} }{\Norm{\vec{v}}} + \frac{\abs{\Norm{\vec{w}} - \Norm{\vec{v}} }}{\Norm{\vec{v}}}.
  \end{align}
  Again, using the triangle inequality, we have $\norm{\vec{v}} = \Norm{\vec{v} - \vec{w} + \vec{w}} \le \Norm{\vec{v} - \vec{w}} + \Norm{\vec{w}}$.
  Thus, $\abs{\Norm{\vec{v}} - \Norm{\vec{w}}} \le \Norm{\vec{v} - \vec{w}} $ holds.
  Therefore, the lemma follows.
}

From this lemma, the error between quantum states
is bounded by the distance between corresponding vectors.
Therefore, we define vectors that describe states $\ket{f}, \ket{f_M}$ and $\ket{\tilde{f}}$,
and derive the upper bound of the distances between those vectors.

Before we define the vectors,
we define vector $\vec{\tilde{x}'}$ that describes state $\ket{\tilde{x}'}$
and vectors $\vec{\tilde{x}}_k \ (k=0,1,\dots,M-1)$.

\begin{defi}
  Let $\epsilon' > 0$.
  For state $\ket{\tilde{x}'}$ such that $\norm{\ket{x'} - \ket{\tilde{x}'} } \le \epsilon'$,
  we define $NM$-dimensional vectors $\vec{\tilde{x}'}$ such that
  $\ket{\tilde{x}'} = \qstateinline{\sum_i \tilde{x}'^{[i]} \ket{i}}$,
  where $\tilde{x}'^{[i]}$ is the $i$-th element of $\vec{\tilde{x}'}$.
  Additionally, we define vectors $\vec{\tilde{x}}_k \in \C^N  (k=0,1,\dots,M-1)$ such that
  $\vec{\tilde{x}'} = (\vec{\tilde{x}}_0^\tenti, \vec{\tilde{x}}_1^\tenti,\dots, \vec{\tilde{x}}_{M-1}^\tenti)^\tenti$.
\end{defi}
Using vectors $\vec{\tilde{x}'}$ and $\vec{\tilde{x}}_k$ defined above,
we define the vectors that describe states $\ket{f}, \ket{f_M}$ and $\ket{\tilde{f}}$.
\begin{defi}\label{def:fs}
  We define
  \begin{align}
    \vec{f} := \frac{M}{\norm{\vec{x'}}}f(A)\vec{b},
    \label{eq:vecf}
  \end{align}
  and
  \begin{align}
    \vec{f}_M := \frac{M}{\norm{\vec{x'}}}f_M(A)\vec{b} = \frac{1}{\norm{\vec{x'}}}\sum_{k=0}^{M-1}g_k\vec{x}_k, \qquad
    \vec{\tilde{f}}_M := \frac{1}{\Norm{\vec{\tilde{x}'}}}\sum_{k=0}^{M-1}\tilde{g}_k\vec{\tilde{x}}_k.
    \label{eq:vecfm}
  \end{align}
\end{defi}
Indeed, using vectors $\vec{f}, \vec{f}_M$ and $\vec{\tilde{f}}_M$ defined as above,
states $\ket{f}, \ket{f_M}$ and $\ket{\tilde{f}}$ can be
described as
\begin{align}
  \ket{f} = \qstate{\sum_{i}^{}{f}^{[i]}\ket{i}}, \quad
  \ket{f_M} = \qstate{\sum_{i}^{} {f}_M^{\ [i]} \ket{i} },
  \quad \text{and}\quad
  \ket{\tilde{f}} = \qstate{ \sum_{i}^{} {\tilde{f}}_M^{\ [i]}\ket{i} },
\end{align}
%---------------------------------------------------------------------------------------------------------------------------------------------
respectively, where $f^{[i]}, f_M^{\ [i]}$, and $\tilde{f}_M^{\ [i]}$ is the $i$-th element of vector $\vec{f}, \vec{f}_M$ and $\vec{\tilde{f}}_M $, respectively.

Next, we derive the upper bound of distances $\norm{\vec{f}- \vec{f}_M}$ and $\norm{ \vec{f}_M - \vec{\tilde{f}}_M}$
in the following lemmas.

\begin{lem}\label{lemma:f-fm}
  For vectors $\vec{f}$ and $\vec{f}_M$, the following equation holds.
  \begin{align}
    \Norm{\vec{f} - \vec{f}_M} \le  \frac{M}{\norm{\vec{x'}}}\Norm{f(A) - f_M(A)}\norm{\vec{b}}.
    \label{eq:f-fm}
  \end{align}
\end{lem}
\proof{
  Given the definitions of vectors $\vec{f}$ and $\vec{f}_M$, we have Eq.~\eqref{eq:f-fm}.
}
\begin{lem}\label{lemma:fm-tildefm}
  For vectors $\vec{f}_M$ and $\vec{\tilde{f}}_M$, the following equation holds.
  \begin{align}
    \Norm{\vec{f}_M - \vec{\tilde{f}}_M }
    \le
    \sqrt{M}B \left(\epsilon' + \frac{r^L}{1-r}\right).
    \label{eq:fm-tildefm}
  \end{align}
\end{lem}
\proof{
  Using the triangle inequality, we obtain
  \begin{align}
    \Norm{\vec{f}_M - \vec{\tilde{f}}_M }
    &\le
    \Norm{
      \vec{f}_M - \frac{1}{\Norm{\vec{\tilde{x}'}}} \sum_{k=0}^{M-1} g_k \vec{\tilde{x}}_k
    }
    +
    \Norm{
      \frac{1}{\Norm{\vec{\tilde{x}'}}} \sum_{k=0}^{M-1} g_k \vec{\tilde{x}}_k - \vec{\tilde{f}}
    }\notag\\
    &=
    \Norm{
      \sum_{k=0}^{M-1}g_k \left(\frac{\vec{x}_k}{\norm{\vec{x'}}} - \frac{\vec{\tilde{x}_k}}{\Norm{\vec{\tilde{x}'}}}\right)
    }
    +
    \Norm{
      \frac{1}{\Norm{\vec{\tilde{x}'}}} \sum_{k=0}^{M-1} (g_k - \tilde{g}_k) \vec{\tilde{x}}_k
    }\notag\\
    &\le
    \sum_{k=0}^{M-1}\abs{g_k}\Norm{ \left(\frac{\vec{x}_k}{\norm{\vec{x'}}} - \frac{\vec{\tilde{x}_k}}{\Norm{\vec{\tilde{x}'}}}\right) }
    +
    \frac{1}{\Norm{\vec{\tilde{x}'}}}\sum_{k=0}^{M-1}\abs{ g_k - \tilde{g}_k } \Norm{ \vec{\tilde{x}}_k }
    .
    \label{bound1}
  \end{align}
  Applying the Cauchy--Schwarz inequality, we have
  \begin{align}
    \Norm{\vec{f}_M - \vec{\tilde{f}}_M }
    &\le
    \left(\sum_{k=0}^{M-1}\abs{g_k}^2\right)^{\frac{1}{2}}
    \left(\sum_{k=0}^{M-1}\Norm{ \left(\frac{\vec{x}_k}{\norm{\vec{x'}}} - \frac{\vec{\tilde{x}_k}}{\Norm{\vec{\tilde{x}'}}}\right) }^2 \right)^{\frac{1}{2}} \notag\\
    &\qquad\qquad +
    \frac{1}{\Norm{\vec{\tilde{x}'}}}\left(\sum_{k=0}^{M-1}\abs{ g_k - \tilde{g}_k }^2 \right)^{\frac{1}{2}}
    \left( \sum_{k=0}^{M-1} \Norm{ \vec{\tilde{x}}_k }^2 \right)^{\frac{1}{2}}.
    \label{eq:ubofrm-tildefm}
  \end{align}
  Recall that the maximum value of $|f(z)|$ on disk $|z| \le R$ is $B$ and $\beta < R$.
  Thus, we have $|g_k| = |f(\beta\e^{\im\theta_k})\e^{\im\theta_k}| \le B$ and
  \begin{align}
    |g_k - \tilde{g_k}|
    = \abs{ f(\beta\e^{\im\theta_k}) - \tilde{f}_L(\beta\e^{\im\theta_k}) } \le \sum_{j=L}^\infty \abs{a_j}\beta^j  \le \frac{Br^L}{1 - r},
  \end{align}
  from Cauchy's estimate $|a_j| \le B/R^j$,
  where $r = \beta/R$.
  Furthermore,
  \begin{align}
    \sum_{k=0}^{M-1}\Norm{ \left(\frac{\vec{x}_k}{\norm{\vec{x'}}} - \frac{\vec{\tilde{x}_k}}{\Norm{\vec{\tilde{x}'}}}\right) }^2
    =   \Norm{ \frac{\vec{x'}}{\norm{\vec{x'}}} - \frac{\vec{\tilde{x}'}}{\Norm{\vec{\tilde{x}'}}}  }^2
    = \Norm{\ket{x'} - \ket{\tilde{x}'}}^2 \le \epsilon'^2
  \end{align}
  and
  \begin{align}
    \sum_{k=0}^{M-1} \Norm{ \vec{\tilde{x}}_k }^2 = \Norm{\vec{\tilde{x}'}}^2
  \end{align}
  hold.
  Thus, we have Eq.~\eqref{eq:fm-tildefm}.
}

From the above lemmas, we see the upper bounds of
$\norm{\vec{f} - \vec{f}_M}$ and $\norm{\vec{f}_M - \vec{\tilde{f}}_M}$.
Combining the above lemmas, Lemmas \ref{normOfAprimeInv}, \ref{lemma:veclemma}, and Proposition \ref{approxTheorem},
we have the following proposition on the error of Algorithm 1.

\begin{proposition}[Error of Algorithm 1]\label{Manddelta}
  Consider the quantum algorithm described in Section \ref{QuantAlgo}.
  We assume that the improved version of the HHL algorithm in Step 1 outputs quantum state $\ket{\tilde{x}'}$
  such that   $\norm{\ket{x'} - \ket{\tilde{x}'}}\le \epsilon'$, where $\epsilon'$ is the positive constant.
  We assume that we used $U_{\tilde{f}_L}$ instead of $U_f$ in Step 2.
  Then, for the error of state $\ket{\tilde{f}}$, which is the actual output state of Algorithm 1,
  the following holds.
  \begin{align}
    \Norm{ \ket{f} - \ket{\tilde{f}} }
    &\le
    \frac{2}{F}\left( \frac{(\beta^{-1})^M}{1 - (\beta^{-1})^{M}} + \frac{r^M}{1 - r^M} + \epsilon' + \frac{r^L}{1-r} \right),
   \label{f-ftilde}
  \end{align}
  where $F = \norm{f(A)\ket{b}}(1 - \beta^{-1})/B$.
\end{proposition}
\proof{
  First, we derive the upper bound of $\norm{\vec{f} - \vec{\tilde{f}}_M}$.
  Using the triangle inequality,
  \begin{align}
    \Norm{\vec{f} - \vec{\tilde{f}}_M}
    &\le \Bigl\lVert \vec{f} - \vec{f}_M \Bigr\rVert + \Norm{\vec{f}_M - \vec{\tilde{f}}_M} \notag\\
    &\le \frac{M\norm{\vec{b}}}{\norm{\vec{x'}}} \Norm{f(A) - f_M(A)} + \sqrt{M}B \left(\epsilon' + \frac{r^L}{1-r}\right),
  \end{align}
  where the second inequality used Eq.~\eqref{eq:f-fm} in Lemma \ref{lemma:f-fm} and Eq.~\eqref{eq:fm-tildefm} in Lemma \ref{lemma:fm-tildefm}.
  From Lemma \ref{lemma:veclemma} and the definition of vector $\vec{f}$,
  we obtain
  \begin{align}
    \Norm{ \ket{f} - \ket{\tilde{f}} }
    &= \Norm{ \frac{\vec{f}}{\Norm{\vec{f}}} - \frac{\vec{\tilde{f}}_M}{\norm{\vec{\tilde{f}}_M}}  } \notag\\
    &\le 2\frac{\Norm{\vec{f} - \vec{\tilde{f}}_M}}{\Norm{\vec{f}}} \notag\\%\label{eq:ub1} \\
    &\le 2\frac{\norm{\vec{x'}}}{M\Norm{f(A)\vec{b}}} \left( \frac{M}{\norm{\vec{x'}}} \Norm{f(A) - f_M(A)}\norm{\vec{b}} + \sqrt{M}B \left(\epsilon' + \frac{r^L}{1-r}\right) \right) \notag\\
    &\le \frac{2}{\norm{f(A)\vec{b}}} \left( \Norm{f(A) - f_M(A)}\norm{\vec{b}} + \frac{\norm{\vec{x'}}B}{\sqrt{M}}\left(\epsilon' + \frac{r^L}{1-r}\right) \right) \notag\\
    &\le \frac{2}{\norm{f(A)\ket{b}}}\left( \Norm{f(A) - f_M(A)} + \frac{B}{1- \beta^{-1}}\left(\epsilon' + \frac{r^L}{1-r}\right) \right), \label{eq:ub2}
  \end{align}
  where the last inequality used $\norm{\vec{x'}} \le \Norm{A'^{-1}}\sqrt{M}\norm{\vec{b}}$
  and $\norm{A'^{-1}} \le (1 - \beta^{-1})^{-1}$ in Lemma \ref{normOfAprimeInv}.
  From Eq.~\eqref{fAminusfMA} in Proposition \ref{approxTheorem} and $\norm{A} \le 1 < \beta < R$,
  we have
  \begin{align}
    \norm{f(A) - f_M(A)} \le \frac{B}{1 - \beta^{-1}}\left( \frac{(\beta^{-1})^M}{1 - (\beta^{-1})^M} + \frac{r^M}{1 - r^M} \right).
  \label{eq:ub3}
  \end{align}
  Thus, by applying this to Eq.~\eqref{eq:ub2}, we have Eq.~\eqref{f-ftilde}.
}

From this proposition, we see an appropriate setting
of parameters $\epsilon', M$ and $L$ to bound the error
$\norm{\ket{f} - \ket{\tilde{f}}}$ by positive constant $\epsilon$.
We describe this in the following corollary.

\begin{cor}\label{cor:parameters}
  Consider the quantum algorithm described in Section \ref{QuantAlgo}.
  We assume that the improved version of the HHL algorithm in Step 1 outputs quantum state $\ket{\tilde{x}'}$
  such that $\norm{\ket{x'} - \ket{\tilde{x}'}}\le \epsilon'$, where $\epsilon'$ is a positive constant.
  We assume that we used $U_{\tilde{f}_L}$ instead of $U_f$ in Step 2.
  Then, to upper bound $\norm{\ket{f} - \ket{\tilde{f}}}$ by positive constant $\epsilon$ such that $0 \le \epsilon \le 1/2$,
  it is sufficient to set parameters $\epsilon', M$ and $L$ satisfying
  \begin{align}
    \epsilon' \le \frac{1}{8}F\epsilon, \ \
     M \ge \max\left\{\frac{1}{1-\beta^{-1}}, \frac{1}{1-r} \right\} \log\left(\frac{8}{F\epsilon} + 1\right), \ \
     L \ge \frac{1}{1-r} \log\left( \frac{8}{(1 - r)F\epsilon} \right),
    \label{eq:parameters}
  \end{align}
  where $F = \Norm{f(A)\ket{b}}(1 - \beta^{-1})/B$.
\end{cor}
\proof{
    If the following inequalities hold,
    then the right side of Eq.~\eqref{f-ftilde} in Proposition \ref{Manddelta}
    is bounded by positive constant $\epsilon$.
    \begin{align}
      \frac{2}{F}\frac{(\beta^{-1})^M}{1 - (\beta^{-1})^{M}} \le \frac{1}{4}\epsilon, \qquad
      \frac{2}{F}\frac{r^M}{1 - r^M} \le \frac{1}{4}\epsilon,  \qquad
      \frac{2}{F}\epsilon' \le \frac{1}{4}\epsilon, \qquad
      \frac{2}{F}\frac{r^L}{1-r} \le \frac{1}{4}\epsilon.
    \label{eq:boundterm}
    \end{align}
    Thus, to upper bound $\norm{\ket{f} - \ket{\tilde{f}}}$ by positive constant $\epsilon$,
    it is sufficient to set $\epsilon', M$ and $L$ satisfying
    \begin{align}
      \epsilon' \le \frac{1}{8}F\epsilon, \ \
       M \ge \max\left\{\frac{1}{\log \beta},  \frac{1}{\log(1/r)} \right\} \log\left(\frac{8}{F\epsilon} + 1\right), \ \
       L \ge \frac{1}{\log(1/r)} \log\left( \frac{8}{(1 - r)F\epsilon} \right).
    \end{align}
    Because $\beta > 1$ and $0 < r < 1$,
    inequalities $1/\log{\beta} < 1/(1-\beta^{-1})$ and $1/\log(1/r) < 1/(1-r)$ hold, respectively.
    Therefore, the corollary follows.
}

\subsection{Success probability}\label{subsec:SuccessProb}
In this section, we show a lower bound of the success probability in Step 4 of Algorithm 1 when there is an error.
Ideal success probability $p$ can be described as
\begin{align}
  p &= \frac{C^2 M \Norm{f_M(A)\vec{b}}^2}{\norm{\vec{x'}}^2} \notag\\
    &= \frac{C^2}{M}\Norm{\vec{f}_M}^2.
\end{align}
Thus, the actual success probability can be described as
\begin{align}
  \tilde{p} &= \frac{\tilde{C}^2}{M}\Norm{\vec{\tilde{f}}_M}^2 \notag\\
            &= \frac{1}{\alpha^2M}\Norm{\vec{\tilde{f}}_M}^2,
\end{align}
where $\alpha = \abs{a_0} + \abs{a_1}\beta + \cdots + \abs{a_{L-1}}\beta^{L-1}$.
We show the lower bound of $\tilde{p}$ in the following proposition.

\begin{proposition}[Success probability of Algorithm 1]\label{prop:successprob}
  Consider the quantum algorithm described in Section \ref{QuantAlgo}.
  We assume that the improved version of the HHL algorithm in Step 1 outputs quantum state $\ket{\tilde{x}'}$
  such that   $\norm{\ket{x'} - \ket{\tilde{x}'}}\le \epsilon'$, where $\epsilon'$ is a positive constant.
  We assume that we used $U_{\tilde{f}_L}$ instead of $U_f$ in Step 2.
  If we set parameters $\epsilon', M$ and $L$ satisfying Eq.~\eqref{eq:parameters}, then
  \begin{align}
    \tilde{p} \ge \left(\frac{3}{4}F(1-r)\right)^2
    \label{eq:tildep}
  \end{align}
  holds, where $F=\norm{f(A)\ket{b}}(1-\beta^{-1})/B$.
\end{proposition}
\myproof{
  First, we consider the lower bound of $\norm{\vec{\tilde{f}}_M}^2$.
  Eq.~\eqref{eq:ub2} in the proof of Proposition \ref{Manddelta}
  is upper bounded by positive constant $\epsilon$ such that $0 \le \epsilon \le 1/2$
  when we set $\epsilon', M$ and $L$ satisfying Eq.~\eqref{eq:parameters} in Corollary \ref{cor:parameters}.
  This implies that $2\norm{\vec{f} - \vec{\tilde{f}}_M}/\norm{\vec{f}} \le \epsilon$.
  Using the triangle inequality and $0 \le \epsilon \le 1/2$, we have
  \begin{align}
    \Norm{\vec{f}}
    \le \norm{\vec{f} - \vec{\tilde{f}}_M } + \norm{\vec{\tilde{f}}_M}
    \le \frac{\epsilon}{2}\norm{\vec{f}} + \norm{\vec{\tilde{f}}_M}
    \le \frac{1}{4}\norm{\vec{f}} + \norm{\vec{\tilde{f}}_M}.
  \end{align}
  Therefore, we have $\norm{\vec{\tilde{f}}_M} \ge (3/4)\norm{\vec{f}}$.
  From $\vec{f} = (M/\norm{\vec{x'}})f(A)\vec{b}$, we have
  \begin{align}
    \Norm{\vec{\tilde{f}}_M}
    \ge \frac{3}{4}\norm{\vec{f}}
    &= \frac{3}{4}\cdot\frac{M}{\norm{\vec{x'}}} \Norm{f(A)\vec{b}} \notag \\
    &\ge \frac{3}{4}\cdot\frac{\sqrt{M}}{\Norm{A'^{-1}}}\Norm{f(A)\ket{b}} \notag \\
    &\ge \frac{3}{4}\sqrt{M}(1-\beta^{-1})\Norm{f(A)\ket{b}},
    \label{eq:lb1}
  \end{align}
  where the second inequality used $\norm{\vec{x'}}\le\Norm{A'^{-1}}\sqrt{M}\norm{\vec{b}}$ and
  the third inequality used $\norm{A'^{-1}} \le (1-\beta^{-1})^{-1}$ in Lemma \ref{normOfAprimeInv}.
  Next, we consider the upper bound of $\alpha$. From Cauchy's estimate $|a_j| \le B/R^j$, we have
  \begin{align}
    \alpha = \abs{a_0} + \abs{a_1}\beta + \cdots + \abs{a_{L-1}}\beta^{L-1}
    \le B \sum_{j=0}^{L-1}r^j \le \frac{B}{1 - r},
  \end{align}
  where $r = \beta/R$.
  Thus, the lower bound of $\tilde{p}$ is
  \begin{align}
    \tilde{p} = \frac{1}{\alpha^2M}\Norm{\vec{\tilde{f}}_M}^2
    \ge \left(\frac{3}{4} (1 - \beta^{-1})\Norm{f(A)\ket{b}}\frac{(1-r)}{B} \right)^2
    = \left(\frac{3}{4}F(1-r)\right)^2.
  \end{align}
}

\subsection{Proof of the main result}\label{subsec:ProofOfTheo}
Finally, we prove our main theorem (Theorem \ref{MainResult}) by combining
Corollary \ref{Runtime}, Corollary \ref{cor:parameters} and Proposition \ref{prop:successprob}.

\noindent
\textbf{Proof of Theorem \ref{MainResult}.}
  We consider the quantum algorithm described in Section \ref{QuantAlgo}.
  We assume that the improved version of the HHL algorithm in Step 1 outputs quantum state $\ket{\tilde{x}'}$
  such that   $\norm{\ket{x'} - \ket{\tilde{x}'}}\le \epsilon'$, where $\epsilon'$ is a positive constant.
  We assume that we used $U_{\tilde{f}_L}$ instead of $U_f$ in Step 2.
  %
  % We describe the complexity of Algorithm 1 when we use the amplitude amplification \cite{BHMT:2002}.
  Then, the query and gate complexity from Step 1 to 3 are
  Eqs.~\eqref{queries} and \eqref{gates}, respectively, in Corollary \ref{Runtime}.
  Before we measure the qubits in Step 4, we use amplitude amplification \cite{BHMT:2002}.
  Then, by repeating Steps 1 to 3 $O(\sqrt{1/\tilde{p}})$ times,
  the success probability increases to a constant, which is close to $1$.
  Therefore, we can obtain $\ket{\tilde{f}}$ using
  \begin{align}
    O\left( \frac{1}{\sqrt{\tilde{p}}}  d\kappa'^2 \log^2 \left( \frac{d\kappa'}{\epsilon'} \right) \right)
    \text{ queries to $\mathcal{P}_A$ and }
    O\left( \frac{1}{\sqrt{\tilde{p}}} \kappa'\log\left( \frac{d\kappa'}{\epsilon'} \right)  \right)
    \text{ uses of $\mathcal{P}_{\vec{b}}$},
  \label{proofTheorem:eq:query}
  \end{align}
  with gate complexity
  \begin{align}
    O\Biggl(
    \frac{1}{\sqrt{\tilde{p}}}
      \biggl\{
      d\kappa'^2 \log^2 \left(  \frac{d\kappa'}{\epsilon'} \right)
      \left[ \log(NM)  +
      \log^{\frac{5}{2}}\left( \frac{d \kappa'}{\epsilon'} \right) \right]
      +
      L + \log(M)\log(L)
      \biggr\}
    \Biggr),
    \label{proofTheorem:eq:gate}
  \end{align}
  where $d$ is the sparsity of matrix $A$ and $\kappa' = 1/(1-\beta^{-1})$.
  From Corollary \ref{cor:parameters}, if we set parameters $\epsilon', M$ and $L$ satisfying
  \begin{align}
    \epsilon' \le \frac{1}{8}F\epsilon, \ \
     M \ge \max\left\{\frac{1}{1-\beta^{-1}}, \frac{1}{1-r} \right\} \log\left(\frac{8}{F\epsilon} + 1\right), \ \
     L \ge \frac{1}{1-r} \log\left( \frac{8}{(1 - r)F\epsilon} \right),
  \end{align}
  then $\norm{\ket{f} - \ket{\tilde{f}}} \le \epsilon$ holds,
  where $F = \norm{f(A)\ket{b}}/(B\kappa')$ and $r = \beta/R$.
  Furthermore, from Proposition \ref{prop:successprob}, $\sqrt{1/\tilde{p}}$ is
  \begin{align}
    \sqrt{\frac{1}{\tilde{p}}} \le \frac{4}{3}\frac{1}{F(1-r)}.
    \label{eq:1/tildep}
  \end{align}
 Thus,
 we can see the complexity of Algorithm 1 when we use the amplitude amplification;
 that is, state $\ket{\tilde{f}}$ such that
  $\norm{\ket{f} - \ket{\tilde{f}}} \le \epsilon$ can be obtained
  \begin{align}
    O\left( \frac{d\kappa'^2}{F(1-r)} \log^2 \left( \frac{d\kappa'}{F}\frac{1}{\epsilon} \right) \right)
    \text{ queries to $\mathcal{P}_A$ and }
    O\left( \frac{\kappa'}{F(1-r)}\log\left( \frac{d\kappa'}{F}\frac{1}{\epsilon} \right)  \right)
    \text{ uses of $\mathcal{P}_{\vec{b}}$},
  \end{align}
  with gate complexity
  \begin{align}
    & O\Biggl(
      \frac{d\kappa'^2}{F(1-r)} \log^2 \left(  \frac{d\kappa'}{F}\frac{1}{\epsilon} \right)
      \left[
      \log(N)  +
      \log(\gamma) +
      \log^{\frac{5}{2}}\left( \frac{d \kappa'}{F}\frac{1}{\epsilon} \right)
      \right]
      \notag \\
    &\qquad\qquad  +
      \frac{1}{F(1-r)^2}\log\left(\frac{1}{F(1-r)}\frac{1}{\epsilon} \right)
      +
      \frac{1}{F(1-r)}\log(\gamma) \log\left(\frac{1}{1-r}\right)
    \Biggr),
  \end{align}
  where $\gamma = \max\{ \kappa', 1/(1-r) \}$.
  \qed

% ------------------------------------------------------------------------------
% \input{v36_Conclusion}
\section{Conclusion}\label{Conclusion}
In this paper, for matrix $A$, vector $\vec{b}$ and complex function $f$,
we proposed a quantum algorithm to compute quantum state
$\ket{f} = f(A)\ket{b}/\norm{f(A)\ket{b}}$, where $\ket{b} = \qstateinline{\sum_{i}b^{[i]}\ket{i}}$ is the quantum state corresponding to
$\vec{b} = (b^{[0]}, b^{[1]},\dots,b^{[N-1]})^\tenti$.
The proposed method used Cauchy's integral formula and the trapezoidal rule
to replace the problem of computing matrix functions
with the problem of solving linear systems.
Using the improved version of the HHL algorithm,
our algorithm output state $\ket{f}$ with $\poly(\log(1/\epsilon))$ runtime.

Our method shows that we can obtain $\ket{f}$ even if $A$ is not Hermitian.
This is because the HHL algorithm is applicable to linear systems even
when the coefficient is not Hermitian.

To use our method,
complex function $f(z)$ must be analytical on the disk with center $0$ and radius $\beta$, where $\beta > \norm{A}$.
For example, our method cannot be applied to the case of $f(z) = z^{-1.5}$.
Therefore, extending our method to be applied to such a case remains future work.

Additionally,
there are various approaches to obtaining quantum state $\ket{f}$
using Cauchy's integral formula and the trapezoidal rule.
Furthermore, the approach of computing the weighted sum of the solutions seems to be applicable to computing other targets.
Solving linear system $(\e^{\im\theta_k}I - A/\beta)\vec{x}_k = \vec{b}$ is related to Krylov subspace methods.
Therefore, investigating the relation between those methods and our quantum algorithm is also future work.

%%%%%%%%%%%%%%%%%%%%%%%%%%%%%%%%%
% \tableofcontents

\nonumsection{Acknowledgements}
\noindent
This work has been supported in part by JSPS KAKENHI Grant Numbers JP16H04367, JP18H05392.
We thank Maxine Garcia, PhD, from Edanz Group (www.edanzediting.com/ac) for editing a draft of this manuscript.
We also thank the editor and the anonymous referee for the comments for the original manuscript of this paper.

% \newpage
\nonumsection{References}
\noindent

%
% References置き場
%
%    \textit{},

\end{document}